\documentclass[%
aip,
amsmath,amssymb,
reprint,%
]{revtex4-1}

\usepackage{graphicx}% Include figure files
\usepackage{dcolumn}% Align table columns on decimal point
\usepackage{bm}% bold math
\usepackage{bm}        % bold math
\usepackage{amsfonts}  % Common math fonts
\usepackage{amsmath}   % Common math functions
\usepackage{amssymb}   % Common math symbols
\usepackage{subfig}
\usepackage{subfloat}  
\usepackage{float}
\usepackage{bm}

\usepackage[utf8]{inputenc}
\usepackage[T1]{fontenc}
\usepackage{mathptmx}
\usepackage{etoolbox}
\usepackage{multirow}

\usepackage[colorlinks=true,
linkcolor=cyan,
citecolor=cyan,
urlcolor=cyan,
]{hyperref}%

\usepackage{xcolor}

\makeatletter
\def\@email#1#2{%
	\endgroup
	\patchcmd{\titleblock@produce}
	{\frontmatter@RRAPformat}
	{\frontmatter@RRAPformat{\produce@RRAP{*#1\href{mailto:#2}{#2}}}\frontmatter@RRAPformat}
	{}{}
}%
\makeatother
\begin{document}
	
	\preprint{AIP/123-QED}
	
	\title{A Robust Data-Driven Model for Flapping Aerodynamics \\ under different hovering kinematics}
	% Force line breaks with \\
	
	\author{Andre Calado}
	\altaffiliation[Now at ]{Department of Mechanical and Aerospace Engineering, The George Washington University.}
	%\email{andre.calado@vki.ac.be}
	\affiliation{
		von Karman Institute for Fluid Dynamics, Waterloosesteenweg 72, Sint-Genesius-Rode, Belgium
	}
	
	\author{Romain Poletti}
	%\email{romain.poletti@vki.ac.be}
	
	\affiliation{
		von Karman Institute for Fluid Dynamics, Waterloosesteenweg 72, Sint-Genesius-Rode, Belgium
	}
	\affiliation{
		Department of Electromechanical, Systems and Metal Engineering, Ghent University,
		Sint-Pietersnieuwstraat 41, Gent, Belgium
	}
	\author{Lilla K. Koloszar}
	\affiliation{
		von Karman Institute for Fluid Dynamics, Waterloosesteenweg 72, Sint-Genesius-Rode, Belgium
	}
	\author{Miguel A. Mendez}
	\affiliation{
		von Karman Institute for Fluid Dynamics, Waterloosesteenweg 72, Sint-Genesius-Rode, Belgium
	}
	\email{mendez@vki.ac.be}

	\date{\today}
	
	\begin{abstract}
		Flapping Wing Micro Air Vehicles (FWMAV) are highly manoeuvrable, bio-inspired drones that can assist in surveys and rescue missions. Flapping wings generate various unsteady lift enhancement mechanisms challenging the derivation of reduced models to predict instantaneous aerodynamic performance. 
		In this work, we propose a robust CFD data-driven, quasi-steady (QS) Reduced Order Model (ROM) to predict the lift and drag coefficients within a flapping cycle. 
		The model is derived for a rigid ellipsoid wing with different parameterized kinematics in hovering conditions. 
		The proposed ROM is built via a two-stage regression.
		The first stage, defined as `in-cycle' (IC), computes the parameters of a regression linking the aerodynamic coefficients to the instantaneous wing state. The second stage, `out-of-cycle' (OOC), links the IC weights to the flapping features that define the flapping motion. 
		The training and test dataset were generated via high-fidelity simulations using the overset method, spanning a wide range of Reynolds numbers and flapping kinematics. The two-stage regressor combines Ridge regression and Gaussian Process (GP) regression to provide estimates of the model uncertainties. 
		The proposed ROM shows accurate aerodynamic predictions for widely varying kinematics. The model performs best for smooth kinematics that generate a stable Leading Edge Vortex (LEV). Remarkably accurate predictions are also observed in dynamic scenarios where the LEV is partially shed, the non-circulatory forces are considerable, and the wing encounters its own wake.
		
	\end{abstract}
	
	\maketitle
	\section{Introduction}
	\label{sec:1}
	
	Micro Aerial Vehicles (MAV) have been a subject of active research since the late '90s, with DARPA (Defense Advanced Research Projects Agency) establishing specific design requirements \cite{mcmichael,hylton}. These small-scale (<15 cm) robots have potential applications for surveillance, rescue missions, or even martian surveys \cite{Phan2019, Pohly2021}.  Bio-inspired Flapping-Wing Micro Air Vehicles (FWMAVs) are more viable than fixed-wing MAVs for stability and agility, since the former cannot hover and rotary wings are generally noisier \cite{Badrya2016}.

	Comparable to insects and small birds, FWMAVs fly at low $Re$ (below $10^4$), defined as $Re= U_r \overline{c}/\nu$, where $U_r$ is a reference velocity, defined in the following section, $\overline{c}$ is the average chord and $\nu$ is the air kinematic viscosity. The low $Re$ results in laminar flows, but the flapping introduces unsteady lift-enhancement mechanisms such as the Leading Edge Vortex (LEV) \cite{Chin2016,ShyyWeiYongshengLianJianTangDragosViieru2008}, rotational circulation and added mass forces \cite{Sane2002}. Considering a wing rigid enough to have negligible deformation during the flapping, the relative importance of these mechanisms depends on the Reynolds number, the reduced frequency $k$, the wing aspect ratio $AR$, and the Rossby number $Ro$\cite{ShyyWeiYongshengLianJianTangDragosViieru2008,Tang2008,balaras,Lang2022,Wang2022,Guo2022}. These quantities are defined precisely in the next section.
	
	Many simplified aerodynamic models have been developed to compute the aerodynamic forces and moments on a flapping wing for various wing kinematics and flight regimes. These models are essential tools for fast predictions, required for real-time model based control or design optimization, and can be classified as steady, quasi-steady and unsteady\cite{Ansari2006, Xuan2020}. 
	
	Steady models are typically based on actuator disk or vortex-based approaches but can only provide time-averaged forces \cite{Xuan2020,Badrya2016}. Quasi-steady models link instantaneous forces to instantaneous states of the wing's kinematics and flow field, thus missing history effects. Yet, these are the most popular approaches because they balance simplicity and accuracy. The simplest QS models cannot account for LEV formation and shedding, typically assuming small pitch angles (uncharacteristic of natural flyers). 
	
	A recent review of aerodynamic models focusing on QS models is given by \citet{Xuan2020}. Spanwise discretization using Blade Element Models (BEM) is typically used to account for wing shape variability, with the total force decomposed in translational, rotational and added mass contributions. Semi-empirical models use coefficients calibrated from experimental or CFD data, but these generalize poorly outside the range of kinematics and flow conditions in the calibration.
	
	Unsteady models introduce functional dependencies on the history of the flapping kinematics and the flow. These are derived from aerodynamic theory and do not rely on the small pitch angle approximation. Nevertheless, typical unsteady models (2D or quasi-3D) have difficulty handling stroke reversal, as the resulting flow separation undermines the validity of the Kutta-Joukowski condition \cite{Ansari2006,Knowles2007}. These models are computationally more expensive but still have difficulties in describing 3D phenomena such as the span-wise motion of the LEV. This is an essential mechanism at low Rossby numbers, such as those encountered in FWMAV applications. 
	
	The limits of the analytical models have pushed research toward developing data-driven Reduced Order Models (ROMs), which aim to be computationally faster and sufficiently accurate for real-time control and optimization of FWMAVs \cite{Zheng2013,Cai2021,Zheng2020}. The development of ROMs typically entails dimensionality reduction from high-fidelity experiments/simulations to derive low-order representations that can adequately describe the aerodynamic performance. When informed or constrained by physics, Machine Learning (ML) and data-driven methods have shown great potential in offering solutions to such class problems \cite{Brunton2020,Vinuesa2021}. Two major trends in data-driven models for flapping aerodynamics have emerged in recent years: state-space models and quasi-steady models built using various regression techniques from machine learning.
	
	State-space models are designed to capture highly transient peaks for arbitrary kinematics \cite{Brunton2013}. These model dynamical systems through latent states from the past to predict future states. Notable examples are the model by \citet{Taha2014}, based on the extension of Duhamel's principle for arbitrary lift curves to capture the LEV effect, and the compact state-space model by \citet{Bayiz2021}, solely based on wing kinematics and calibrated on experimental data. Modern variants of this class of methods are the models based on Volterra series, which describe causal, time-invariant, non-linear systems with fading memory\cite{Liu2017,Ruiz2022}.
	
	Considering quasi-steady ROMs, empirical models have been derived by \citet{Nakata2015} from the least-squares fitting of coefficients for the different force terms (translation, rotation, and added mass).  \citet{Lee2016} proposed to extend the flexibility of these models by including $Re$, $Ro$, $AR$ and taper ratio (tip chord to root chord) in the empirical coefficients. Model errors up to 20\% were attributed to wing geometry effects, shedding of LEV and wing-wake interaction. \citet{Zheng2020} proposed a data-driven and self-adaptive model combined with an optimizer that searches for the kinematic parameters required to achieve a specific lift coefficient. \citet{Cai2021} derived a CFD data-driven aerodynamic QS model (CDAM) for a bumblebee in a range of forward flight conditions, including a simple aerodynamic model for the moving body. This model is based on semi-empirical laws for the various contributions to the aerodynamic force, closed by five empirical coefficients calibrated on CFD data. 
	
	Within the data-driven approaches, Artificial neural network (ANN) models have also been used to obtain the time-averaged lift coefficient by  \citet{Pohly2021}. The input layer of the ANN consisted of 3 dimensionless parameters: $k$, $Re$, and $AR$, which are the varied parameters in the CFD database for a total of 125 simulations, 25 of which are used as test data.
	
	Even though several QS models have been developed from CFD data, most involve tuning empirical coefficients tied to classical formulations of the aerodynamical mechanisms under stationary conditions and are thus unable to capture relevant dynamics effects such as the wake capture \cite{Xuan2020,Dickinson1999,Hu2020}. Moreover, these models have been developed for narrow kinematic ranges, either sinusoidal or modeled after insects, and do not consider more dynamic motions with broad variation \cite{Nakata2015,Zheng2020,Hu2020,Cai2021}.

	The objective of this study is to leverage machine learning methods to derive a data-driven QS model valid for different wing kinematics, using high-fidelity CFD data for both model training and validation. The proposed approach relies on a two-stage regression combining Ridge regression with Gaussian Process (GP) regression. Essential definitions and model scope are established in section \ref{sec:2}, along with the range of conditions that were explored for the model development.
	Section \ref{sec:methodology} details the setup of the high-fidelity CFD used to generate the training and test data, followed by a description of the proposed regression strategy for deriving our data-driven aerodynamic model. Section \ref{sec:results} collects the results, with a discussion on the model performance and the underlying physics from CFD  visualization. Conclusions and future extensions are outlined in section \ref{sec:conclusions}.

	\section{Definitions and model scope}\label{sec:2}
	
\begin{figure}[h]
		\centering
		\includegraphics[width=0.95\linewidth]{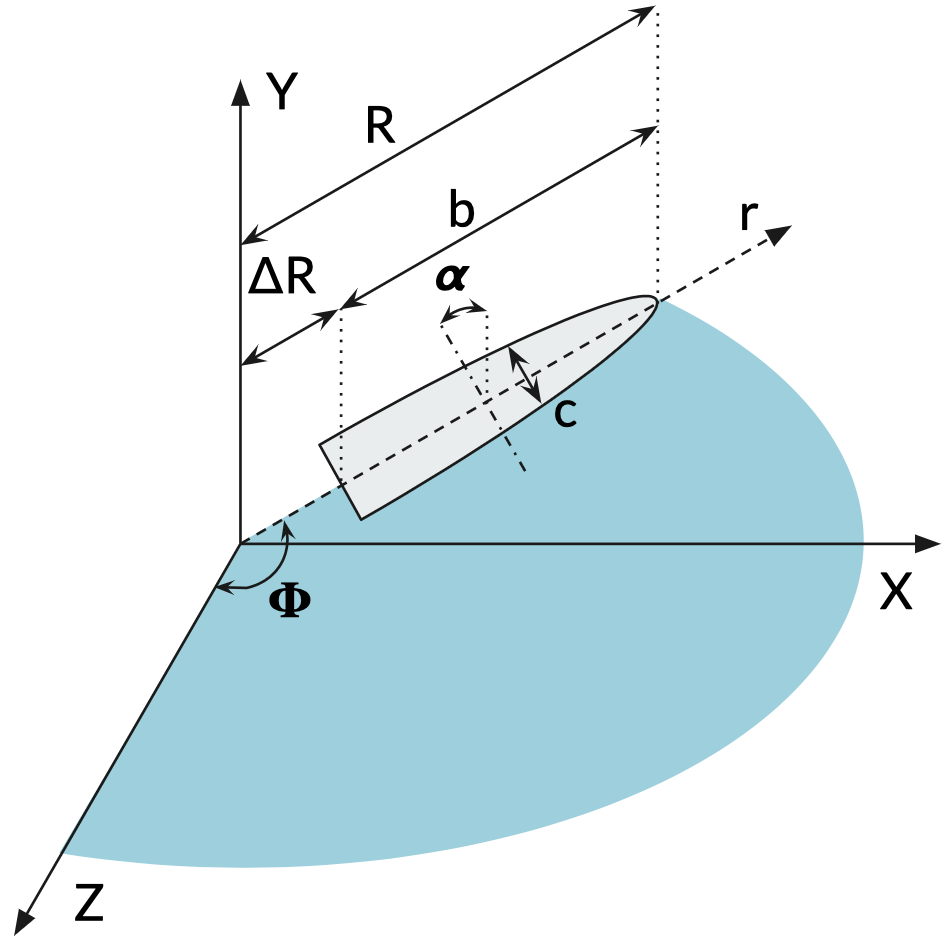}
		\caption{Schematic illustration of the relevant parameters in the flapping of an ellipsoid wing. The full stroke amplitude is $\Phi$, and the instantaneous pitching angle is $\alpha$. $r$ is the local spanwise coordinate from the flapping axis, $\Delta R$ the offset of the wing root, with span $b$ and chord $c$.}
		\label{fig:euler}
	\end{figure}

	The kinematics of a flapping wing is described by three motion angles (shown schematically in Fig. \ref{fig:euler}): the flapping/translation angle ($\phi$) along the stroke plane, the pitching angle (or feathering/rotation angle, $\alpha$), and the elevation angle ($\theta$) of the stroke plane.  This study focuses on hovering conditions and we fix the elevation angle to $\theta=0$° since this angle is known to be of secondary importance in force generation\cite{Bayiz_2018,Sane2001,Qin2014}.

	For hovering flight, the characteristic Reynolds number $Re$ is defined from the average chord $\Bar{c}$, and a reference speed $U_{r}$ defined by the distance spanned by the radius of second moment of area $R_2 = \sqrt{\frac{1}{S}\int_0^R cr^2 dr}$ over a flapping period \cite{ShyyWeiYongshengLianJianTangDragosViieru2008,Pohly2021}, where $r$ is the local spanwise coordinate (cf. Fig. \ref{fig:euler}). At $\theta=0$°, the Reynolds number is defined as:
	\begin{equation}
		\label{eq:reynolds}
		Re = \frac{U_{r} \Bar{c}  }{\nu}  =  \frac{ 2 f \Phi R_2  \Bar{c} }{\nu} 
	\end{equation}
	where $\Phi$ is the full stroke flapping amplitude and $\nu$ the kinematic viscosity of air.
	The reduced frequency for hovering flight is defined as: 
	\begin{equation}
		\label{eq:k_hover}
		k =\frac{\pi f \Bar{c}}{U_r} = \frac{\pi}{2 \Phi Ro}
	\end{equation}
	%where $AR= b/\Bar{c}$ is the wing's aspect ratio. 
	where the Rossby number is defined as $Ro = R_2 / \Bar{c}$ and is kept constant in this study.

	We consider the same thin rigid semi-elliptical wing as in \citet{Lee2016}, with pitching axis centered at mid-chord. The span is 50 mm with $AR=b/\Bar{c}=3.25$, $Ro=2.97$ and a thickness equal to 1\% of span.  The wing motion is defined from the popular parametrization of \citet{Berman2007}, with no pitching offset or phase difference. The time-varying flapping and pitching angles are given respectively as:
	\begin{eqnarray}
		\label{eq:phi}
		\phi (t) = \frac{A_{\phi}}{\arcsin (K_{\phi})}\arcsin [K_{\phi}\sin(2 \pi  ft)] \\
		\label{eq:alpha}
		\alpha (t) = \frac{A_{\alpha}}{\tanh (K_{\alpha})}\tanh [K_{\alpha}\sin(2 \pi  ft)] \,.
	\end{eqnarray}
	Thus the wing motion has a total of five independent parameters that describe the kinematics: the amplitudes ($A_{\phi},A_{\alpha}$) and shape factors ($K_\phi,K_\alpha$) for both the flapping and pitching angles, and the flapping frequency $f$.
	The parameter bounds in the current study are defined in Table \ref{tab:parameter-range}.
	
	\begin{table}[!ht]
		\caption{Kinematic parameter range.}
		\begin{tabular}{lc}
			Parameter \qquad \qquad &  Range \\ \hline
			$Re$ &  $10^2$ - $10^4 $   \\
			$A_{\phi} (=\Phi / 2) $ &  15$^{\circ}$ - 75$^{\circ}$      \\
			$A_{\alpha}$ &  15$^{\circ}$ - 75$^{\circ}$    \\
			$K_{\phi}$ &  0.01 - 0.99   \\
			$K_{\alpha}$ & 0.01 - 10 
		\end{tabular}
		\label{tab:parameter-range}
	\end{table}
	
	Forces acting on the wing are represented in terms of lift and drag coefficients, with forces normalized by the wing area ($S$) and dynamic pressure using the reference velocity, that is:
	
	\begin{eqnarray}
		\label{eq:cl}
		C_L = \frac{L}{0.5 \rho U_{r}^2 S}  \\
		\label{eq:cd}
		C_D = \frac{D}{0.5 \rho U_{r}^2 S} \,.
	\end{eqnarray}
	
	The lift force $L$ defined as the vertical component (along Y), and drag force $D$ defined as the component on the stroke plane (XZ), being positive if opposing the flapping velocity.
	
	\section{Methodology}\label{sec:methodology}
	\subsection{CFD Model}\label{sec:cfd-model}
	
	High-fidelity CFD simulations were used to generate the database of aerodynamic forces for the model training and testing. The simulations were carried out with the finite volume CFD code OpenFOAM\textregistered  (v2012). The simulations are 3D, incompressible, unsteady, and use the overset mesh technique (\textit{overPimpleDyMFoam} solver). The overset method has been used by different authors\cite{Liu2009,Nakata2015,Cai2021,Ruiz2022} to study the dynamics of wing motion and the reader is referred to \citet{Hadzic2005} for more information on its working principle.

\begin{figure}[!h]
		\centering
		\subfloat[Component mesh.]{\label{fig:overset}\includegraphics[width=0.75\linewidth]{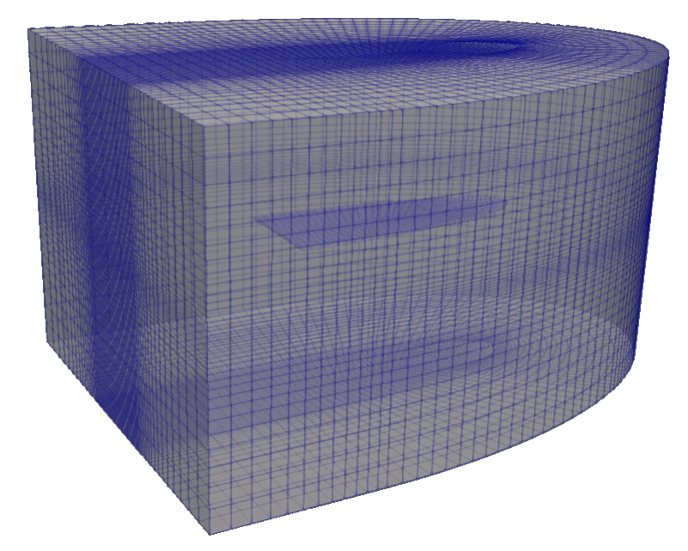}} \qquad
		\subfloat[Background and component meshes combined.]{\label{fig:background}\includegraphics[width=0.75\linewidth]{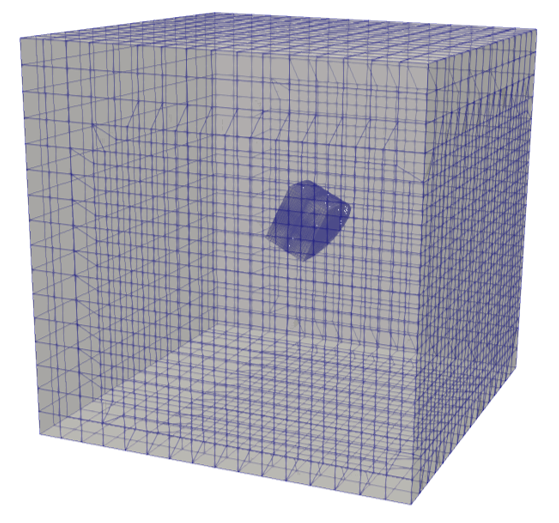}}
		\caption{Overset mesh used in CFD.}
		\label{fig:meshes}
	\end{figure}

	For the present work, a structured component mesh is fitted to the wing. The component mesh spans two chord lengths normal from the surface in all directions (similar to \citet{Liu2009}) with inflating cell size, shown in Fig. \ref{fig:overset}. 
	The background grid is a cube with boundaries at least 10 chords away from the wing to avoid any potential boundary effects \cite{Sun2002}. The background grid is locally refined along the wing path and wake to match the cell sizes at the interface of the background and component grids. Figure \ref{fig:background} shows the combined background and component meshes.
	The boundary conditions on the background grid consisted of a zero gauge pressure on all faces, and zero gradient for velocity. The initial conditions assume fluid at rest. To ensure that the extracted lift and drag profiles are representative (periodic), all simulations are computed for five flapping periods, with only the last cycle used for post-processing as usual in the literature\cite{Liu2009,Cai2021,Ruiz2022}.
	
	Regarding numerical schemes, the backward second-order implicit scheme is used for the time discretization, with variable time-stepping and a max Courant number of 1. 
	A second-order linear Gauss scheme is used for spatial discretization, with a limiter for divergence terms. The pre-conditioned conjugate gradient (PCG) iterative solver is used for the cell displacement, the pre-conditioned bi-stable conjugate gradient (Pre-BiCG) solver for pressure, and the symmetric Gauss-Seidel for velocity. The PIMPLE algorithm was employed for the pressure-velocity coupling. 
	A comparison was also made between a laminar flow model (no turbulence modeling) and the dynamic turbulent kinetic energy (TKE) Large Eddy Simulation (LES) turbulence model \cite{LES} at the upper bound of Reynolds for a case with $Re \approx 10^4$. This is considered to be the limit $Re$ above which turbulence influence the aerodynamics and stability of vortical structures\cite{Chin2016}. Nevertheless, the difference between laminar and LES models was found to be negligible. Although laminar models have been used reliably up to this value of $Re$ by various authors\cite{Liu2009,Nakata2015,Lee2016,Cai2021}, the dynamic TKE-equation sub-grid model was  chosen here because of its ability to better adapt to different $Re$ and flow conditions, since the upper bound for $Re$ is near the transition threshold.

	A grid sensitivity study was carried out on the component cell size, evaluating the convergence of the mean lift and drag magnitude for an intermediate $Re \approx 4000$ and smooth harmonic motion ($K_{\phi}=K_{\alpha}=0.01$), with $A_\phi = 60^{\circ}$ and $A_\alpha = 45^{\circ}$. Table \ref{tab:grid_sensitivity} collects the results of this study, with the time-averaged lift ($\langle C_L\rangle$) and drag magnitude ($\langle |C_D|\rangle$) coefficients obtained for three mesh refinements (coarse, medium, and fine).
	A resolution of 130$\times 10^3$ cells in the component domain (medium level) has a deviation of less than 1\% in $\langle  C_L \rangle$, and 3\% in $\langle  |C_D| \rangle$, with respect to the finest grid. Therefore, the intermediate mesh was considered sufficiently accurate.  
	
	\begin{table}[!htb]
		\caption{Grid sensitivity study.}
		\begin{tabular}{lccc}
			Grid  \quad & CG cells [$10^3$] \quad & $\langle  C_L \rangle$ \quad &  $\langle|C_D| \rangle$ \\ \hline
			Coarse & 41                                   & 2.019  & 2.813    \\
			Medium & 130                                  & 2.042  & 2.797    \\
			Fine   & 430                                  & 2.044  & 2.736   
		\end{tabular}
		\label{tab:grid_sensitivity}
	\end{table}
	
\begin{figure}[!ht]
		\centering
		\subfloat[Lift.]{\label{fig:validation-lift}\includegraphics[width=0.8\linewidth]{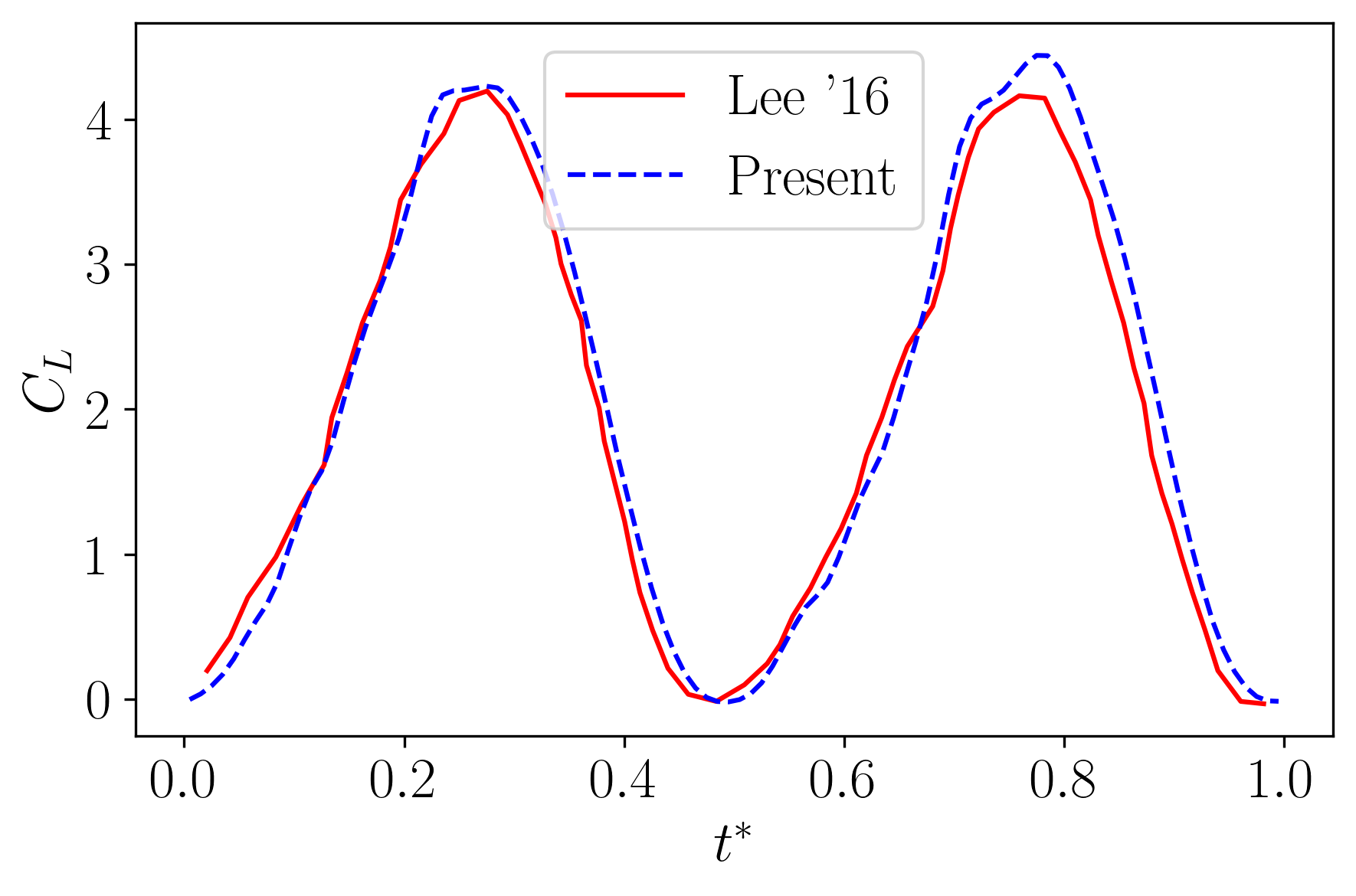}}\quad
		\subfloat[Drag (magnitude).]{\label{fig:validation-drag}\includegraphics[width=0.8\linewidth]{Figure_3a.png}}\\
		\caption{CFD validation against \citet{Lee2016}. Kinematic parameters: $Re \approx 4000, A_\phi = 60^{\circ}$, $A_\alpha = 45^{\circ}, K_{\phi}=K_{\alpha}=0.01$.}
		\label{fig:validation}
	\end{figure}

	The numerical setup is validated against the results of \citet{Lee2016} with same kinematics used in the grid sensitivity study. The agreement is very good with a relative root mean square error below 2\% for both lift and drag profiles as seen in Fig. \ref{fig:validation}. The setup is thus considered adequate for producing the high fidelity data.  A total of 165 simulations are performed. Of these, a 15\% test size (25 cases) were randomly taken as \emph{testing} set; the results in these conditions were not used to train the model but to check its generalization performance. The simulated cases in CFD are defined using Latin Hypercube Sampling (LHS\cite{osti_5571631}), with a specified minimum Euclidean distance between points.

	\subsection{Reduced Order Modeling}\label{sec:rom}

	The proposed ROM is built from a two-stage regression. The first, referred to as `in-cycle' (IC) regression, maps the aerodynamic coefficients to the instantaneous states of the wing kinematics. The second, referred to as `out-of-cycle' (OOC), maps the parameters in the IC regression to flapping conditions.

	The IC regression follows the QS approach of \citet{Bayiz2021} of expressing lift and drag coefficients as a linear combination of nonlinear features.
	Denoting as $C(t)$ the general aerodynamic coefficient (i.e. $C_D$ or $C_L$) at time $t$, we thus have:
	
	\begin{equation}
		\label{lin}
		C(t)=\sum^{n_f}_{j=1} b_j (t) w_j=\mathbf{b}^{\top}(t)\mathbf{w}\,,
	\end{equation} where $\mathbf{b}(t)=[b_1,\dots, b_{n_f}](t)\in\mathbb{R}^{n_f}$ is the vector collecting the $n_f$ features at time $t$, $\mathbf{w}\in\mathbb{R}^{n_f}$ is the vector of parameters (weights) and $^{\top}$ denotes transposition. When relevant, we shall use $\mathbf{w}^L$ and $\mathbf{w}^D$ for the weights linked to the prediction of lift and drag coefficients respectively.
	
	The features selected in this work are the same for both coefficients and reads:
	
	\begin{equation}
		\begin{split}
			\label{Basis}
			\mathbf{b}(t)=[& \cos(\phi), |\ddot{\alpha}|, \dot{\phi}^2, \dot{\alpha}^2, \dot{\phi}\ddot{\phi},\\ & \dot{\alpha}\ddot{\alpha}, \sin(2\alpha)\dot{\alpha}, \sin(2\alpha)\sin(\phi) ](t)\,.
		\end{split}
	\end{equation} 
	
	Thus we have $n_f=8$.
	
	The feature selection was inspired by \citet{Bayiz2021}, and previous QS models in literature \cite{Cai2021, Lee2016, Nakata2015, Zheng2020, Sane2002}, then heuristically improved by trial and error. In particular, the cross terms $\dot{\phi}\ddot{\phi},\dot{\alpha}\ddot{\alpha}, \sin(2\alpha)\dot{\alpha}, \sin(2\alpha)\sin(\phi)$, unconventional in the literature of flapping wings, were found to significantly improve the ROM's accuracy. It is worth noticing that the analytically prescribed wing kinematics (eqs. \ref{eq:phi} and \ref{eq:alpha}) allows computing all features beforehand and link them to the associated kinematic parameters $A_{\phi}, A_{\alpha}, K_{\phi}, K_{\alpha}$. Similarly, the time-averaged features can also be analytically computed. Finally, to give all features equal importance, these have been scaled in the range $[0,1]$ using the maximum value observed within a flapping cycle; the scaling quantities can also be computed analytically from \eqref{eq:phi} and \eqref{eq:alpha}. {Accordingly, the time derivatives in \eqref{Basis} are taken with respect to the non-dimensional time $t^* = t f$; this is equivalent to scaling velocity and accelerations by $f$ and $f^2$ respectively.}
	
	In the machine learning terminology, the identification of the weights $\mathbf{w}$ from a set of data (in this case provided by CFD) is referred to as \emph{training} and was carried out using Ridge regression \cite{Hoerl1970}. Given a set of $n_t$ samples, collected at times $t_*=[t_1,\dots t_{n_t}]$ and $\mathbf{c}(t_*)=[C(t_1),\dots C(t_{n_t})]$ the corresponding lift or drag coefficients, and given $\mathbf{B}(t_*)=[b_1(t_*),\dots b_{n_f}(t_*)]\in\mathbb{R}^{n_t\times 8}$ the matrix collecting the corresponding normalized features along each column, the optimal weights are those that minimize the following cost function
	
	\begin{equation}
		\label{eq:ridge}
		\textrm{min} \quad J(\mathbf{w}) = ||\mathbf{c}(t_*) - \mathbf{B}(t_*) \mathbf{w}||_2 ^2 + \lambda ||\mathbf{w}||_2^2\,,
	\end{equation} where $\lambda$ is a regularizing penalty and $||\bullet||_2$ denotes the $l_2$ norm of a vector. Besides providing better accuracy, the $l_2$ penalty (Tikhonov regularization) was chosen over the $l_1$ penalty (Lasso regression) because the weights showed a large variance over the different range of kinematics, and the sparser model promoted by the $l_1$ regularization would not consistently eliminate the same terms.
	
	The regularization $\lambda$ is the first of the 17 hyper-parameters of the proposed ROM and introduced in this section. 
	
	The solution to the minimization \eqref{eq:ridge} is:
	
	\begin{equation}
		\label{IC_REG}
		\mathbf{w}^*=\bigl(\mathbf{B}^{\top}(t_*)\mathbf{B}(t_*)+\lambda \mathbf{I}_8\bigr)^{-1}\mathbf{B}^{\top}(t_*)\mathbf{c}(t_*)\,
	\end{equation} where $\mathbf{I}_8$ is the $8\times8$ identity matrix. 
	
	These weights are linked to kinematic input parameters in the OOC regression.
	Defining $\mathbf{x}=[Re,k,A_{\alpha},K_{\phi},K_{\alpha}]^{\top}$ as the out-of-cycle flapping parameters (also scaled in $[0,1]$), the OOC regression seeks to identify the mapping $\mathbf{x} \rightarrow \mathbf{w}$, hence $\mathbb{R}^{5}\rightarrow \mathbb{R} ^8$. This was carried out using multivariate Gaussian Process Regression \cite{CarlEdward}, because of its flexibility, sample efficiency and natural formulation of the model uncertainty.
	
	Gaussian processes are probabilistic models that provide the probability distribution over possible functions compatible with observed data. The primary assumption is that any finite sample of these functions is jointly Gaussian distributed. Therefore, given $\mathbf{X}=[\mathbf{x}_1,\cdots\mathbf{x}_{n_p}]\in\mathbb{R}^{5\times n_p}$ a set of possible kinematic parameters, all candidate functions can be sampled using $5$ dimensional Gaussian distribution, here denoted as: 
	
	\begin{equation}
		\label{Prior}
		\mathbf{w}^*_j(\mathbf{X})\sim \mathcal{N}({\mu}_j , \mathbf{K}_j)\,,
	\end{equation} where $\mathbf{w}^*_j$ is the set of values taken by the j-th weight in the model \eqref{lin} for the set of kinematic parameters $\mathbf{X}$, $\bm{\mu}_j=[\mu_j(\mathbf{x}_1),\cdots \mu_j(\mathbf{x}_{n_p})]\in \mathbb{R}^{n_p}$ is the vector of average predictions of the j-th weight for each set of kinematic parameters and $\mathbf{K}_j\in\mathbb{R}^{n_p\times n_p}$ is the covariance matrix for the j-th weight. In this work, the multivariate prediction is constructed by a set of univariate predictions, each having its independent Gaussian process.
	We thus have $8$ processes in $\mathbb{R}^{5}$.
	
	Equation \eqref{Prior} is the \emph{prior} distribution for each weight. We take $\mu_j=0$ for all weights and covariances defined by Gaussian kernels with entries $\mathbf{K}_j[{l,m}]=\kappa_j(\mathbf{x}_l,\mathbf{x}_m)$, with
	
	\begin{equation}
		\label{eq:kernel}
		\kappa_j (\mathbf{x}_l, \mathbf{x}_m) = \textrm{exp}\left(-\frac{||\mathbf{x}_l-\mathbf{x}_m||_2^2}{2\gamma_j^2}\right),
	\end{equation} and $\gamma_j$ the length scale for the j-th process. The eight length scales $\gamma_j$ are hyper-parameters of the surrogate model.  
	
	Given a set of training points $\mathbf{w}^*_j \in\mathbb{R}^{n^*}$ for each of the j-th weights in \eqref{lin}, calibrated by the Ridge regression on kinematic parameters $\mathbf{X}^{*}=[\mathbf{x}_1,\cdots\mathbf{x}_{n^{*}}]\in\mathbb{R}^{5\times n^*}$, the underlying assumption in the out-of-cycle Gaussian Process regression is that the weights $\mathbf{w}'_j \in\mathbb{R}^{n'}$ associated to any set of kinematic parameters  $\mathbf{X}'=[\mathbf{x}_1,\cdots\mathbf{x}_{n'}]\in\mathbb{R}^{5\times n'}$ are joint Gaussian distributed:
	
	\begin{equation}
		\label{eq:Posterior}
		\begin{bmatrix}\mathbf{w}^{*}_j(\mathbf{X}^{*})  \\ \mathbf{w}'_j(\mathbf{X}') \end{bmatrix}\sim \mathcal{N}\Biggl (\begin{bmatrix}\mathbf{0}  \\ \mathbf{0} \end{bmatrix},  \begin{bmatrix}\mathbf{K}_j+\sigma^2_{w_j}\mathbf{I}_{n^*} & \mathbf{K}'_j \\\mathbf{K}_j'^{\top} & \mathbf{K}''_j \end{bmatrix} \Biggr)
	\end{equation} where $\mathbf{K}_j=\kappa_j(\mathbf{X}^*,\mathbf{X}^*)\in\mathbb{R}^{n^*\times n^*}$, $\mathbf{K}'_j=\kappa_j(\mathbf{X}^{*},\mathbf{X}^{'})\in\mathbb{R}^{n^{*}\times n'}$ and  $\mathbf{K}''_j=\kappa_j(\mathbf{X}',\mathbf{X}')\in\mathbb{R}^{n'\times n'}$. Therefore, the probability density function associated to $\mathbf{w}'_j(\mathbf{X}^{'})$ can be obtained using standard rules of conditioning, leading to a multivariate Gaussian distribution of the form:
	
	\begin{equation}
		\label{GPR}
		\mathbf{w}'_j(\mathbf{X}')\sim \mathcal{N}(\bm{\mu}_{\mathbf{w},j}\,,\, \bm{\Sigma_{w,j}})\,
	\end{equation} with
	
	\begin{subequations}
		\begin{equation}
			\bm{\mu}_{\mathbf{w},j}=\mathbf{K'}_j^{\top}\bigl(\mathbf{K}_j+\sigma^2_{w_j}\mathbf{I}_{n^*}\bigr)^{-1}\mathbf{w}^*_j(\mathbf{\mathbf{X}^{*}})\in\mathbb{R}^{n'}
		\end{equation}
		\begin{equation}
			\bm{\Sigma_{w,j}}=\mathbf{K}_j''-\mathbf{K}_j'^{\top}
			\bigl(\mathbf{K}_j+\sigma^2_{w_j}\mathbf{I}_{n'}\bigr)^{-1}\mathbf{K}'_{j}+\sigma^2_{w_j}\mathbf{I}_{n'}\in\mathbb{R}^{n'\times n'}\,.
		\end{equation}
	\end{subequations}
	
	The regularization terms $\sigma^2_{w_j}$ in the inversion of the covariance matrices $\mathbf{K}_j$ are hyper-parameters and are linked to the assumption that an independent and identically distributed (i.i.d.) Gaussian noise is added to the training data. This noise term is introduced to facilitate the pairing of the OOC regression with the IC regression. Together with $\gamma_j$, these parameters control the ability of the GP regression to follow sharp variations in the training data, which in this context could be due to over-fitting in the IC regression. 
	
	The covariance matrices $\bm{\Sigma_{w,j}}$ can be used to provide an uncertainty in the prediction of the weights $\mathbf{w}_j$. For each of $n'$ samples, the entry along the diagonal $\bm{\Sigma}_{w,j}$ provides the posterior variance on the j-th weights. Therefore, for each new sample point it is possible to collect the 8 posterior variances in a diagonal matrix $\bm{\Sigma}_w\in\mathbb{R}^{8\times8}$ and propagate this to the aerodynamic coefficients. The linearity of the model in \eqref{lin} allows to propagate these variances easily to give
	
	\begin{equation}
		\label{eq:variance}
		\bm{\Sigma_{c}} = \mathbf{B}  \bm{\Sigma}_w  \mathbf{B^{\top}} + \sigma_{c}^2\mathbf{I}_{n_t}\,\in\mathbb{R}^{n_t\times n_t},
	\end{equation} where the term $\sigma_{c}^2$ computed is the global variance obtained from the IC regression, that is from the first term of eq. \eqref{eq:ridge}. The diagonal entries in $\bm{\Sigma_{c}}$ provide the expected variance in the corresponding aerodynamic coefficient over all the samples in the time domain.
	
	The ROM model calibration thus depends on seventeen hyper-parameters: the regularization $\lambda$ in \eqref{eq:ridge}, the eight length scales $\gamma_j$ in the kernels in \eqref{eq:kernel} and the eight regularizing variances $\sigma_{w_j}$ in \eqref{eq:Posterior}. In this study, $\lambda$ was fixed to a value of 0.5 which provided a good compromise between overall model accuracy and variance of $\mathbf{w}$.

	The GP hyper-parameters were identified via Hyper-parameter Optimization (HPO), i.e. using an optimizer to minimize the squared $l_2$ error over the dataset. This was combined with K-fold cross-validation, with 10 folds, to minimize overfitting\cite{Fi2010}. 
	Therefore, the model training is repeated 10 times, using at each time 1/10 of the data as validation. Given $\mathbf{c}$ the full set of lift or drag coefficients collected in the training dataset, regardless of their dependence on time and the OOC parameters, and given $\tilde{\mathbf{c}}_f$ the associated predictions of the model calibrated using the $f$-th fold as validation, the cost function to minimize is the average $l_2$ error  
	
	\begin{equation}
		\label{eq:cost_fun}
		\textrm{min} \quad G(\sigma_{w_j}, \gamma_j) = \frac{1}{10}\sum^{10}_{f=1} ||\mathbf{c}- \tilde{\mathbf{c}}_f(\sigma_{w_j}, \gamma_j)||_2^2.
	\end{equation}
	
	This minimization was constrained to the following bounds: $\sigma^2_{w_j} \in [10^{-6} , 10^{-3}]$, $\gamma_j \in [10^{-3} , 1]$. The results of the HPO for the prediction of drag and lift coefficients are collected in Table \ref{tab:hpo}. 
	
	\begin{table}[!htb]
		\caption{Optimized ROM hyper-parameters.}
		\begin{tabular}{lcc}
			Hyper-parameter  \quad &  $  C_L $ \quad &  $C_D $ \\ \hline
			$\sigma^2_{w_1}$ , $\gamma_1$  &   3.51e-04 , 8.21e-01  \qquad  & 1.48e-04 , 7.17e-01   \\
			$\sigma^2_{w_2}$ , $\gamma_2$  &   1.00e-06  , 7.70e-01  \qquad  & 3.73e-04 ,   7.85e-01   \\
			$\sigma^2_{w_3}$ , $\gamma_3$  &  6.07e-05  ,  8.07e-01  \qquad  &   4.34e-04 ,   8.51e-01  \\
			$\sigma^2_{w_4}$ , $\gamma_4$  &   1.24e-05 ,  2.77e-01  \qquad  & 8.66e-04  ,   8.79e-01  \\
			$\sigma^2_{w_5}$ , $\gamma_5$  &  5.06e-05  ,   5.50e-01  \qquad  & 9.55e-04  ,   9.98e-01   \\
			$\sigma^2_{w_6}$ , $\gamma_6$  &  3.92e-04   ,  8.44e-01 \qquad  & 9.22e-04   ,  9.59e-01   \\
			$\sigma^2_{w_7}$ , $\gamma_7$  &  4.37e-05  ,  5.48e-01\qquad  & 9.74e-04 ,  9.44e-01  \\
			$\sigma^2_{w_8}$ , $\gamma_8$  & 4.75e-05  ,  6.14e-01  \qquad  & 9.85e-04  ,   9.86e-01  \\
		\end{tabular}
		\label{tab:hpo}
	\end{table}

	\section{Results and Discussion}\label{sec:results}
	
	This section is divided into two parts. First, we discuss the model performances in terms of global statistics, focusing on the time-averaged predictions (Sec. \ref{4p1}). Then, in Sec. \ref{4p2}, we report on the model performances in predicting instantaneous aerodynamic forces within a flapping cycle. 
	
	\subsection{In-Cycle Average ROM Performance}\label{4p1}
	
\begin{figure*}[!ht]
		\centering
		\includegraphics[width=1\textwidth]{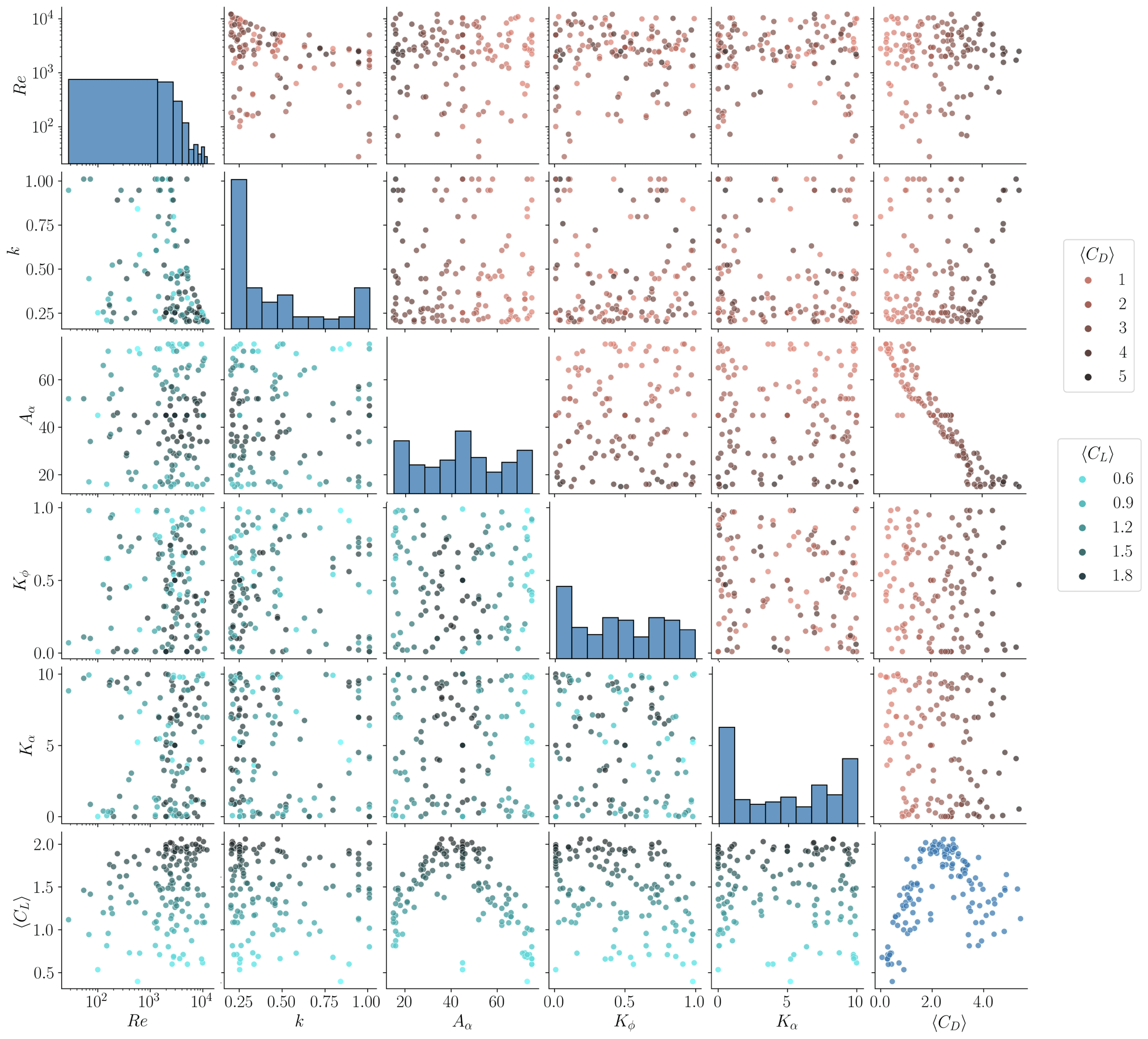}
		\caption{ $\langle C_L \rangle$ (lower left) and $\langle C_D \rangle$ (upper right) from CFD versus OOC parameters. Hue  is proportional to the magnitude of the aerodynamic coefficient. Diagonal plots are histograms of each OOC parameter.} 
		\label{fig:pairwise-mean}
	\end{figure*}

	Figure \ref{fig:pairwise-mean} shows the in-cycle averaged lift $\langle C_L \rangle$ and drag coefficients $\langle C_D \rangle$ from CFD as a function of the out-of-cycle parameters using a grid of scatter plots.
	The plots below the diagonal of the grid are related to $\langle C_L \rangle$ and those above the diagonal are related to $\langle C_D \rangle$. The markers in the scatter plots are coloured by the magnitude of the corresponding coefficient (see legend) to map the points from one figure to the other. The histograms along the diagonal show the (marginal) distributions for each parameter.
	
	These plots give an overview of the density and uniformity of the sampled conditions in the parameter space and illustrate the relative importance of each parameter. All planes involving $K_\alpha,K_\phi,A_\alpha$ are almost uniformly sampled although the plane $k-Re$ is not; this is due to the definition of the sampling boundaries in terms of dimensional variables $f$ and $\Phi$ (table \ref{tab:parameter-range}) and the link between $k$ and $Re$ (equation \ref{eq:reynolds} and \ref{eq:k_hover}). Future work will aim at extending the database on this plane, yet the collected data allowed the training of a robust ROM and revealed unexpected trends. These are qualitatively analyzed in the following.
	
	The most sensitive parameter is the amplitude of the pitching angle $A_{\alpha}$. The drag coefficient $\langle C_D \rangle$ is almost inversely proportional to $A_{\alpha}$, with net contributions from drag and thrust (negative drag) approaching zero for maximum pitching. The lift coefficient $\langle C_L \rangle$ reaches a maximum at $A_{\alpha}\approx$ 45°. For both trends, the minor role of the other parameters is revealed by the sampling in the planes $A_{\alpha}-k$ or $A_\alpha-K_\phi$ and $A_\alpha-K_\alpha$. Similar results on the relation between $A_\alpha$ and the aerodynamic coefficients
	have been documented in the literature\cite{Bos2013,Taha2014,Dickinson1999}. 
	
	The other parameters have much more inter-winded and less expected trends. The plot $\langle C_L\rangle$ vs $Re$ shows a moderate influence of the second on the first, especially at $Re > 10^3$, where the LEV implies a higher suction pressure on the wing upper surface \cite{bhat2019uncoupling}. Considering the points with largest $\langle C_L\rangle$ (hence with  $A_{\alpha}\approx$ 45°) the lift increases from $1.9$ to $2$ when increasing the Reynolds from $10^3$ to $10^4$. This is qualitatively in line with the translational lift dependency $\langle C_L\rangle\propto 1.966-3.94Re^{-0.429}$ reported by \citet{Lee2016}, which was derived at a constant angle of attack and a fixed translational motion in steady conditions, such that the LEV always remains attached to the wing.

	Figure \ref{fig:pairwise-mean} also highlights the dependency between $\langle C_D\rangle$ and $Re$. At moderate $A_{\alpha}$, $\langle C_D\rangle$ increases with $Re$ because of the stronger circulation of the LEV. This trend is less pronounced than QS model predictions\cite{Lee2016} since the shape factors modulate the drag concurrently with $Re$. A higher $K_{\phi}$ and a higher $K_{\alpha}$ both tend to decrease the drag\cite{bhat2020effects}. 
	At the lowest pitching angles $A_{\alpha}\approx 20$°, which also corresponds to the lowest flapping angle $A_{\phi}\approx 20$°, the drag shows a maximum at $\langle C_D\rangle\approx 3.8$ for $Re\approx 10^3$ and decreases to $\langle C_D\rangle\approx 3$ for $Re\approx 10^4$. 
	This trend is not predicted by the relations reported in \citet{Lee2016} since the investigated flapping kinematics strongly differ from the `fixed translational', `arrested rotation', or `continuous rotation' motions they have used to isolate the various contributions of the aerodynamic forces.
	
	Finally, the trends allow identifying the region in the parameter space leading to the best lift-to-drag ratio, which for the analyzed configuration is $\approx 1$, with both coefficients $\approx 2$ (see plot $\langle C_L\rangle$ vs $\langle C_D\rangle$). This optimal region is located at $A_{\alpha}\approx 45$°, large $Re\approx 10^4$ and small $k\approx 0.25$ and is mildly sensitive to $K_\phi$ and $K_\alpha$.
	As we shall see shortly, this region of the parameter space produces accelerations that challenge the quasi-steady assumption underlying most simplified models.
	
	\begin{figure}[!htb]
		\centering
		\subfloat[Lift RMSE box plots.]{\label{fig:box-lift}\includegraphics[width=0.75\linewidth]{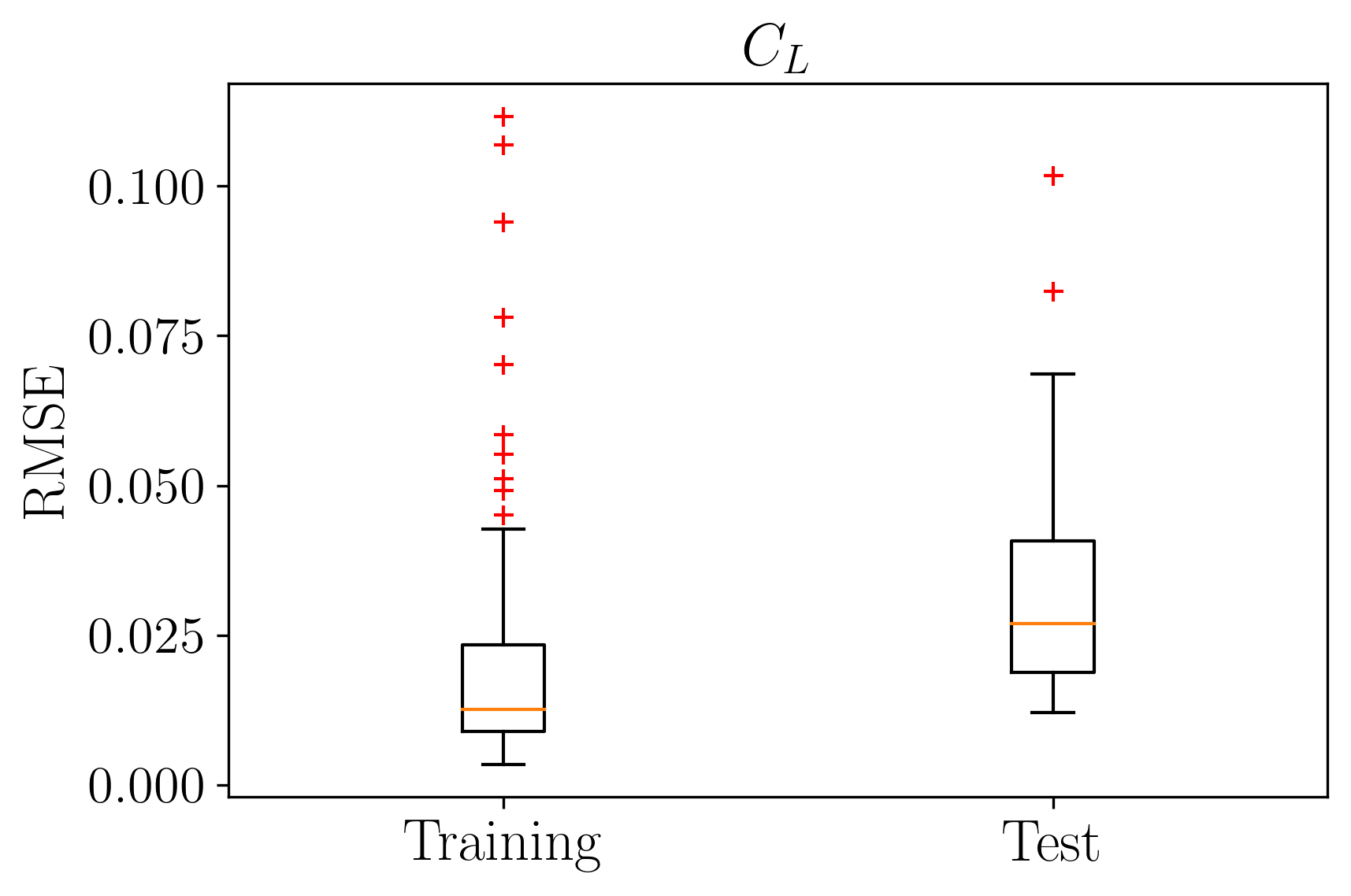}}\\   
		\subfloat[Drag RMSE box plots.]{\label{fig:box-drag}\includegraphics[width=0.75\linewidth]{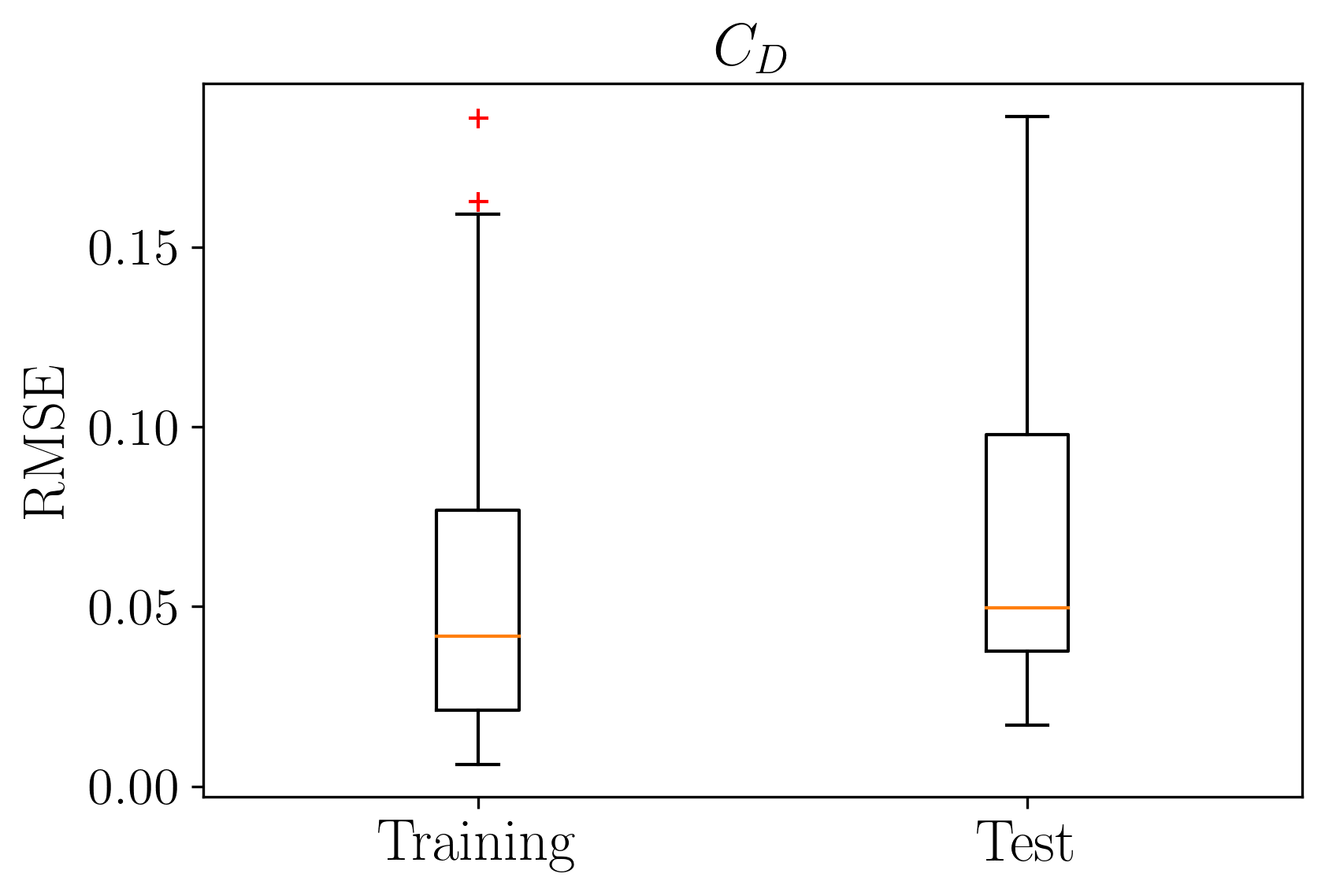}} \\
		\caption{Statistics of the model performance. Figures (a) and (b) show the box plots for the RMSE on the training and the test data for the in-cycle average lift and drag coefficients. } \label{fig:box-error}
	\end{figure}
	
	We now move to analyse the ROM performances in predicting the CFD data, first considering the overall performances and then focusing on the optimal time averaged lift-to-drag ratio region of the parameter space. The evaluation is carried out in terms of Root Mean Squared Error (RMSE) over a flapping cycle. Denoting as $\mathbf{c},\mathbf{\tilde{c}}\in\mathbb{R}^{n_t}$ the vectors collecting the CFD data and the ROM prediction for the $n_t = 100$ samples in a cycle, the RMSE is defined as 
	
	\begin{equation}
		\label{eq:mse}
		RMSE = \frac{1}{n_t}  || \,  \mathbf{c} - \mathbf{\tilde{c}} \, ||_2  \,.
	\end{equation} 
	
	Figures \ref{fig:box-lift} and \ref{fig:box-drag} show the RMSE in both training and test data for lift and drag coefficients. The median error is slightly larger in the test data than in the train data, but their values are satisfactorily small overall. Although more outliers are present in the predictions of the lift coefficient, the spread of the error and the upper quartiles are higher for the drag coefficient. This suggests that the choice of a common basis \eqref{Basis} for both coefficients is sub-optimal, and potential improvements could be achieved by using different bases per coefficient.
	
	\begin{figure}[!htb]
		\centering
		\subfloat[Lift RMSE.]{\label{fig:error-lift}\includegraphics[width=0.75\linewidth]{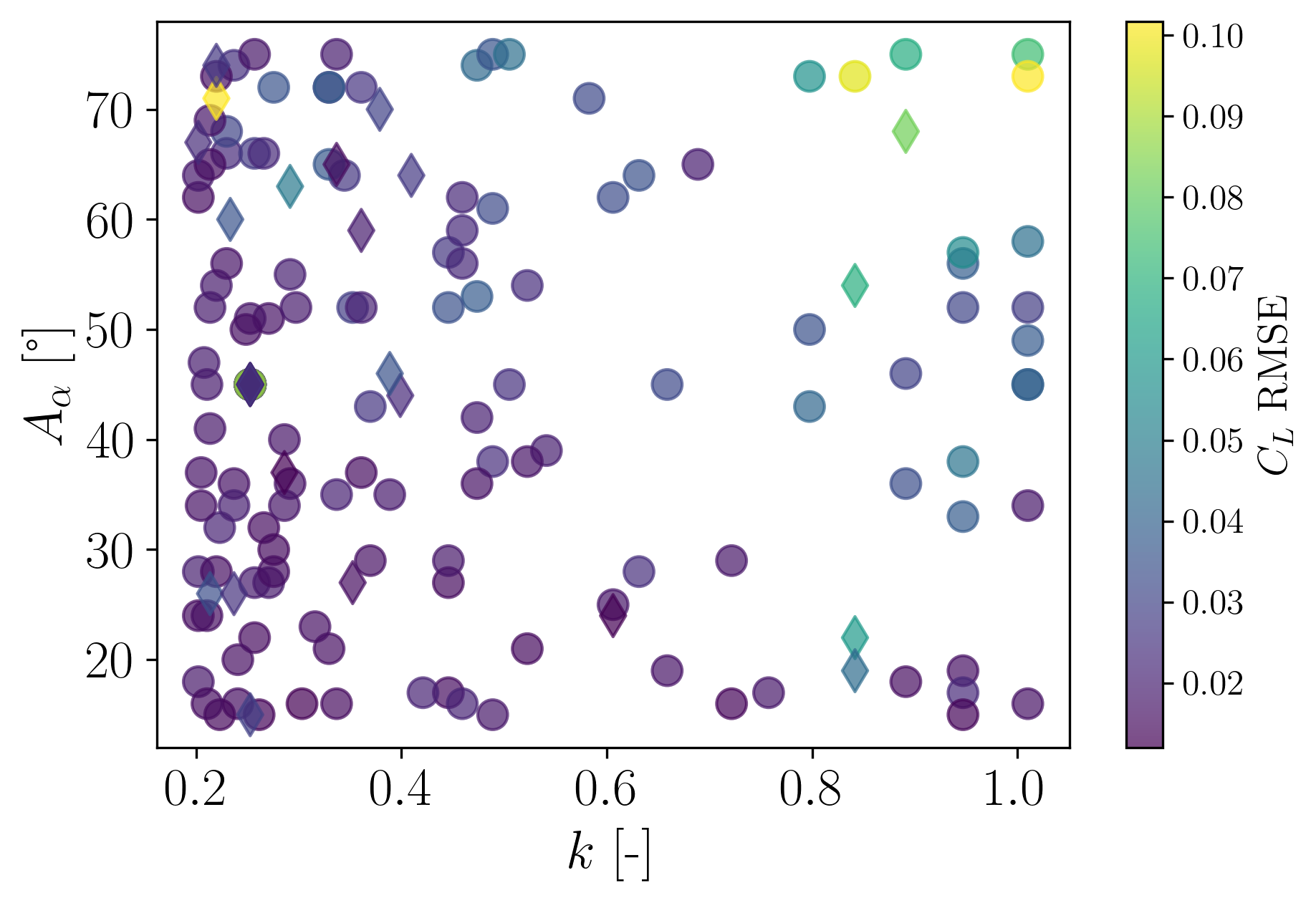}} \\
		\subfloat[Drag RMSE.]{\label{fig:error-drag}\includegraphics[width=0.75\linewidth]{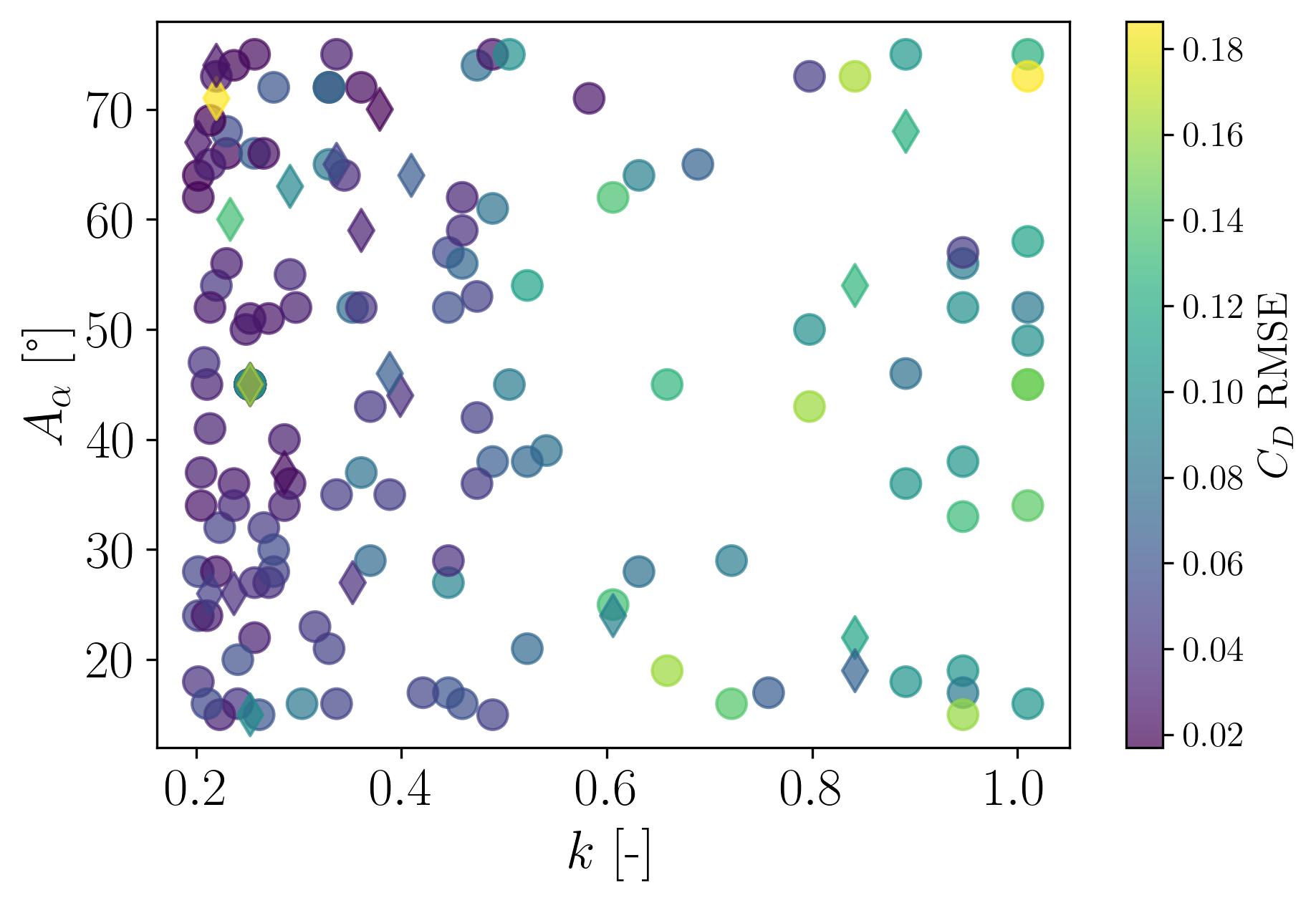}}
		\caption{Dependency of RMSE on $k$ and $A_{\alpha}$. Training and test points are shown with circle and diamond markers respectively.}
		\label{fig:model-error}
	\end{figure}

	\begin{figure*}[!htb]
		\centering
		\subfloat[Lift correlation scatter.]{\label{fig:PCC-CL}\includegraphics[width=0.4\linewidth]{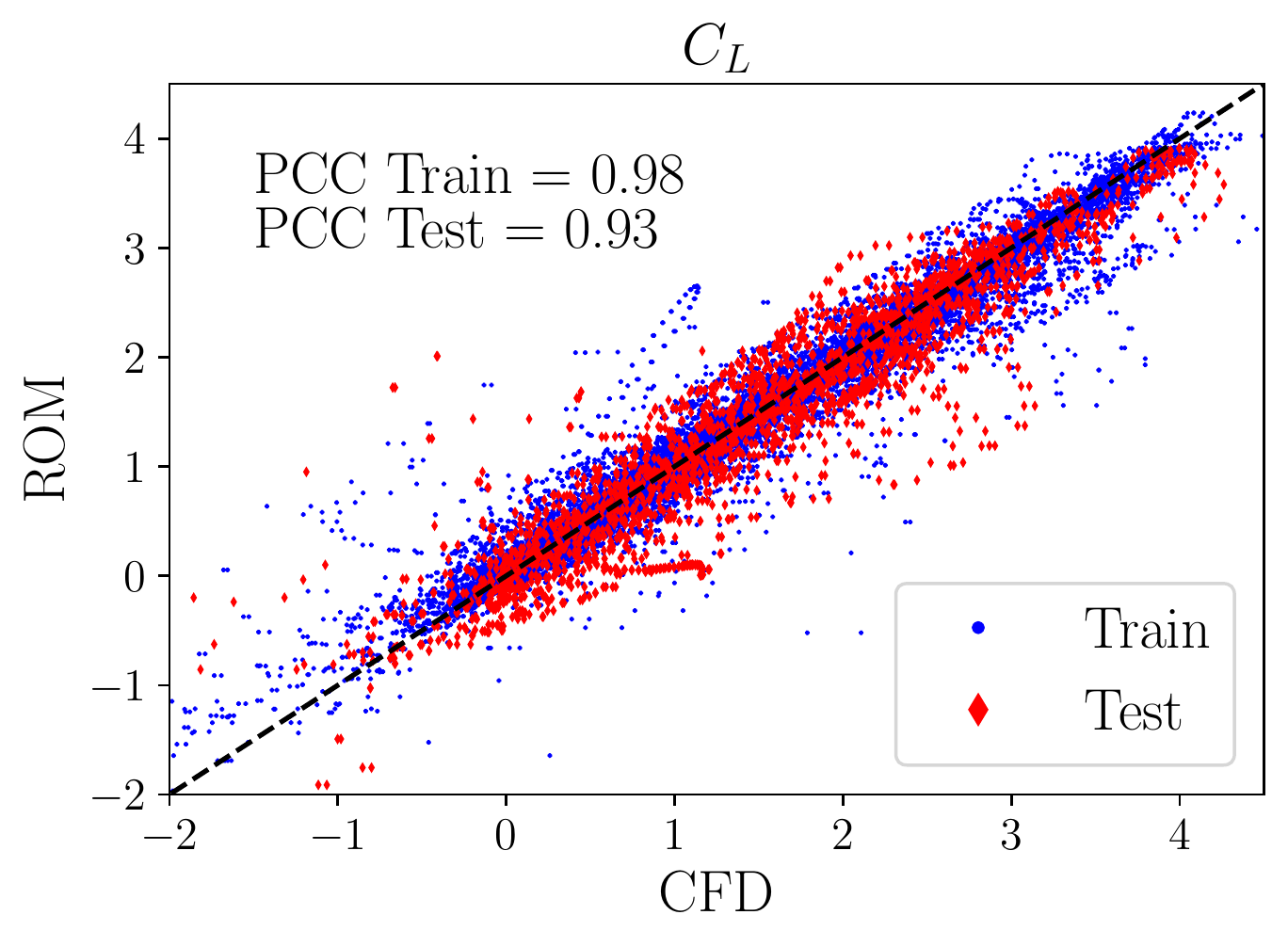}}  \qquad\subfloat[Drag correlation scatter.]{\label{fig:PCC-CD}\includegraphics[width=0.4\linewidth]{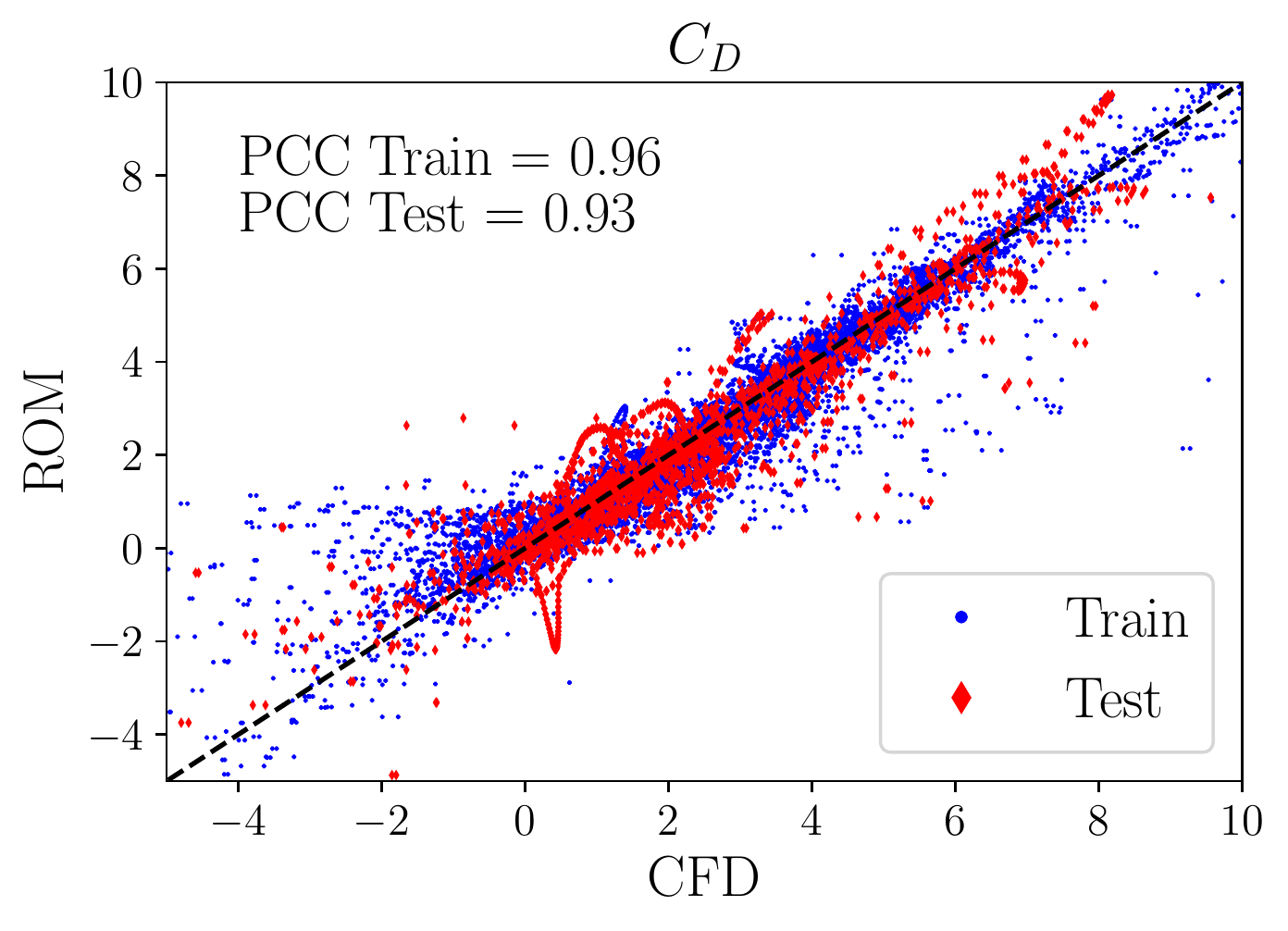}}
		\caption{Statistics of model performance. Figures (a) and (b) compare the ROM and CFD prediction on the training and the test data for the instantaneous ($n_t$ points per simulation) lift and drag coefficients} 
		\label{fig:model_performance}
	\end{figure*}
	
	The time-averaged model performances are further analyzed on the plane $A_{\alpha}-k$ in Figure \ref{fig:model-error} for both lift and drag coefficients. The training data is identified with circle markers, and the test data is identified with diamond markers. As shown in figure \ref{fig:pairwise-mean}, this is the plane explaining the largest portion of the variance in both aerodynamic coefficients. In the area with the best aerodynamic performances, i.e. $k \in [0.2, 0.4]$ and $A_\alpha \in [20^{\circ},60^{\circ}]$, the RMSE is of the order of $0.02$ for the $C_L$ and $0.04$ for the $C_D$. This leads to an RMSE to mean prediction ratio of $\approx 1\%$ for both coefficients.

	\begin{figure*}[!ht]
		\centering
		\includegraphics[width=0.98\textwidth]{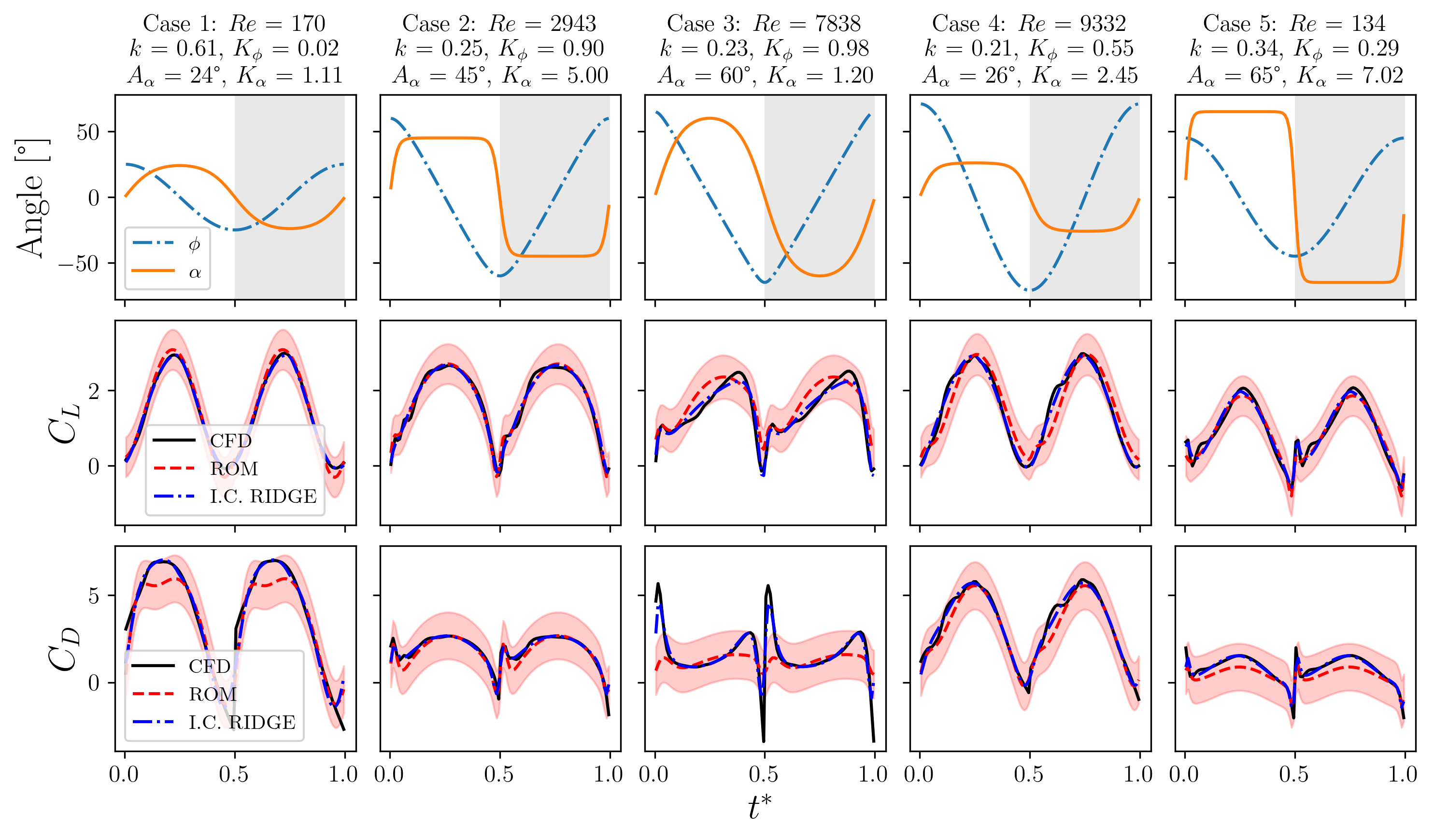}
		\caption{Comparison between CFD, IC Ridge and full ROM (IC+OOC) predictions for 5 different test cases. Top row: wing motion angles. Middle row: $C_L(t)$. Bottom row: $C_D(t)$. The red shaded bands identify the 95\% confidence interval around mean the prediction.} 
		\label{fig:test-cases}
	\end{figure*}
	
	\subsection{In-Cycle Predictions}\label{4p2}
	
	To analyze the overall performances in the instantaneous prediction, we consider the Pearson Correlation Coefficient (PCC) between the CFD data $\mathbf{c}$ and the ROM prediction $\tilde{\mathbf{c}}$:
	
	\begin{equation}
		\label{eq:pcc}
		PCC = \frac{\textrm{Cov}(\mathbf{c},\mathbf{\tilde{c}})}{\sigma(\mathbf{c})\sigma(\mathbf{\tilde{c}})}\,
	\end{equation}
	where $\textrm{Cov}$ and $\sigma$ the covariance and standard deviation operators respectively. The values of PCC are shown in Fig. \ref{fig:PCC-CL} and \ref{fig:PCC-CD} along with the scatter plot of the training and test data, for all the 165 simulations and all the $n_t$ evaluations in a flapping cycle. The PCCs of 0.98 for lift and 0.96 for drag coefficients confirm the quality of the regression with slightly worse performances on the drag. The drop in performances in the test data is acceptable, considering the overall small size of the dataset and the complexity of the function being regressed. The region of larger discrepancy coincides with negative aerodynamic coefficients linked to stroke reversal and wing-wake interaction. 
	
	The in-cycle performances of the ROM are showcased in Figure \ref{fig:test-cases}, which compares the CFD data (continuous black line), and the ROM prediction (dashed red line), i.e. IC + OOC regressions, over a flapping cycle in five representative test cases with largely different kinematics. These are validation test cases not included in the model's training. For each, the first row of plots shows the flapping kinematics, and the figure title recalls the associated parameters. The second and third rows of plots show the instantaneous lift and drag coefficients with the confidence interval around the ROM's prediction. Moreover, to further analyze the strengths and limitations of the proposed approach, each figure also shows the prediction of the IC Ridge regression (blue dashed-point) \eqref{lin} with optimal weights from \eqref{IC_REG}.
	
	\begin{figure*}[!ht]
		\centering
		\includegraphics[width=1\textwidth]{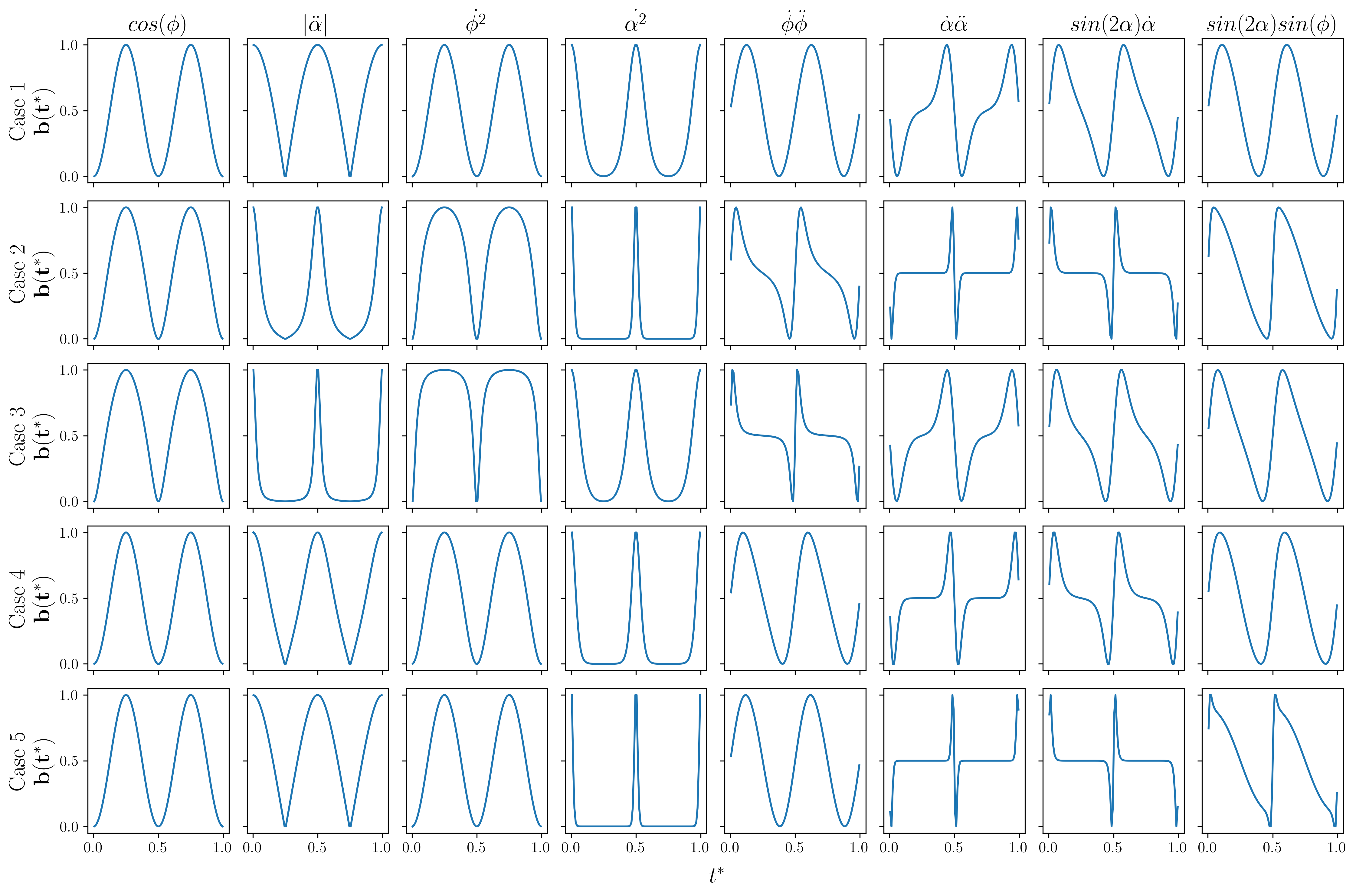}
		\caption{Normalized basis functions for the test cases shown in Fig. \ref{fig:test-cases}. Without normalization, the first four basis elements have a non-zero time average; therefore, these are eligible to model translational contributions. Conversely, the last four have zero time averages and can be used to model rotational and added mass effects. }
		\label{fig:test-cases-basis}
	\end{figure*}

	Cases 1 and 5 have low $Re$, with very different flapping kinematics, with case 5 having a more aggressive pitching motion (higher $K_{\alpha}$). Cases 2 and 4 have $A_{\phi} > A_{\alpha}$, with similar $k$ but different pitching kinematics and Reynolds number. In these four cases, the ROM predictions of both aerodynamic coefficients are in excellent agreement with the CFD data. These test cases highlight the versatility of the ROM model under different wing kinematics, particularly up to moderate values of $A_{\alpha}$. Moreover, as the CFD is overall within the confidence intervals, these test cases also illustrate the reliable prediction of the model uncertainties. On the other hand, the model hits its limits on test case 3, where a relatively high pitching amplitude combined with high $Re$ and $K_{\phi}$ lead to sharp transients at each half-stroke.

	\begin{figure*}[!ht]
		\centering
		\includegraphics[width=0.69\textwidth]{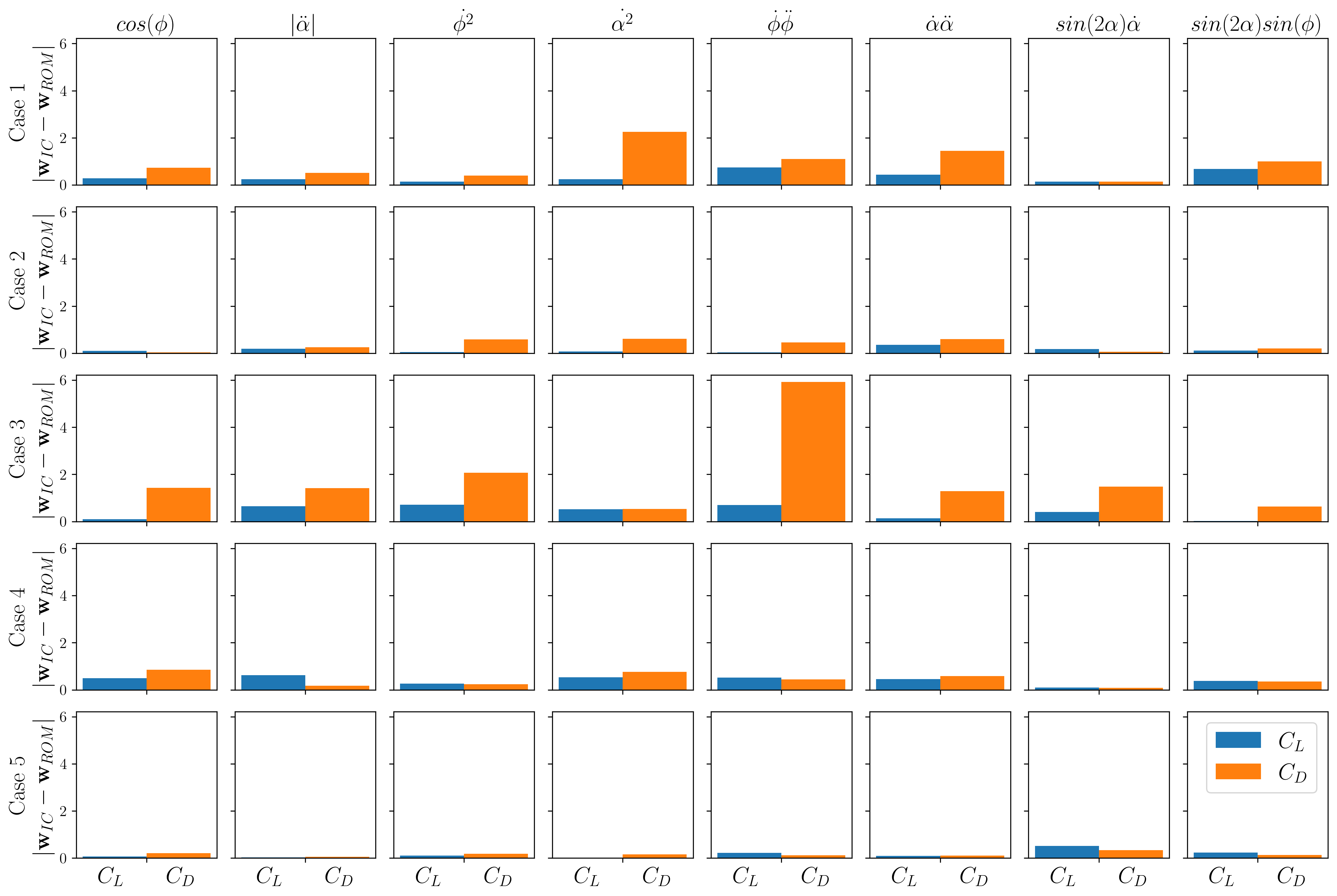}
		\caption{Absolute error between predicted weights from ROM and IC Ridge weights for the test cases shown in Fig. \ref{fig:test-cases}. }
		\label{fig:test-cases-weight-error}
	\end{figure*}

	\begin{figure*}[!htb]
		\centering
		\subfloat[$\mathbf{w}^L$ vs. $\mathbf{x}$.]{\label{fig:wL-ooc}\includegraphics[width=0.88\linewidth]{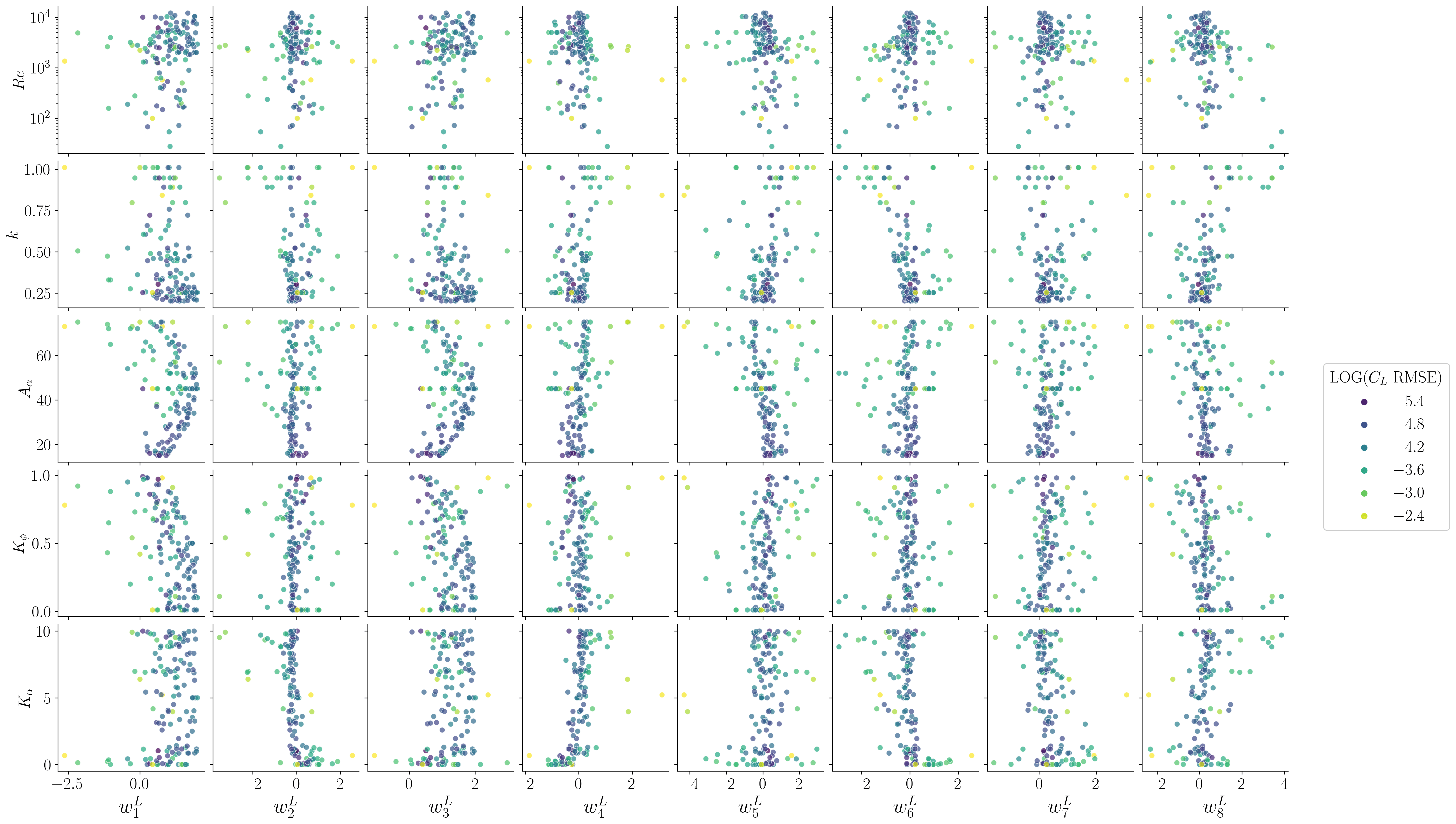}}\quad 
		\subfloat[$\mathbf{w}^D$ vs. $\mathbf{x}$.]{\label{fig:wD-ooc}\includegraphics[width=0.9\linewidth]{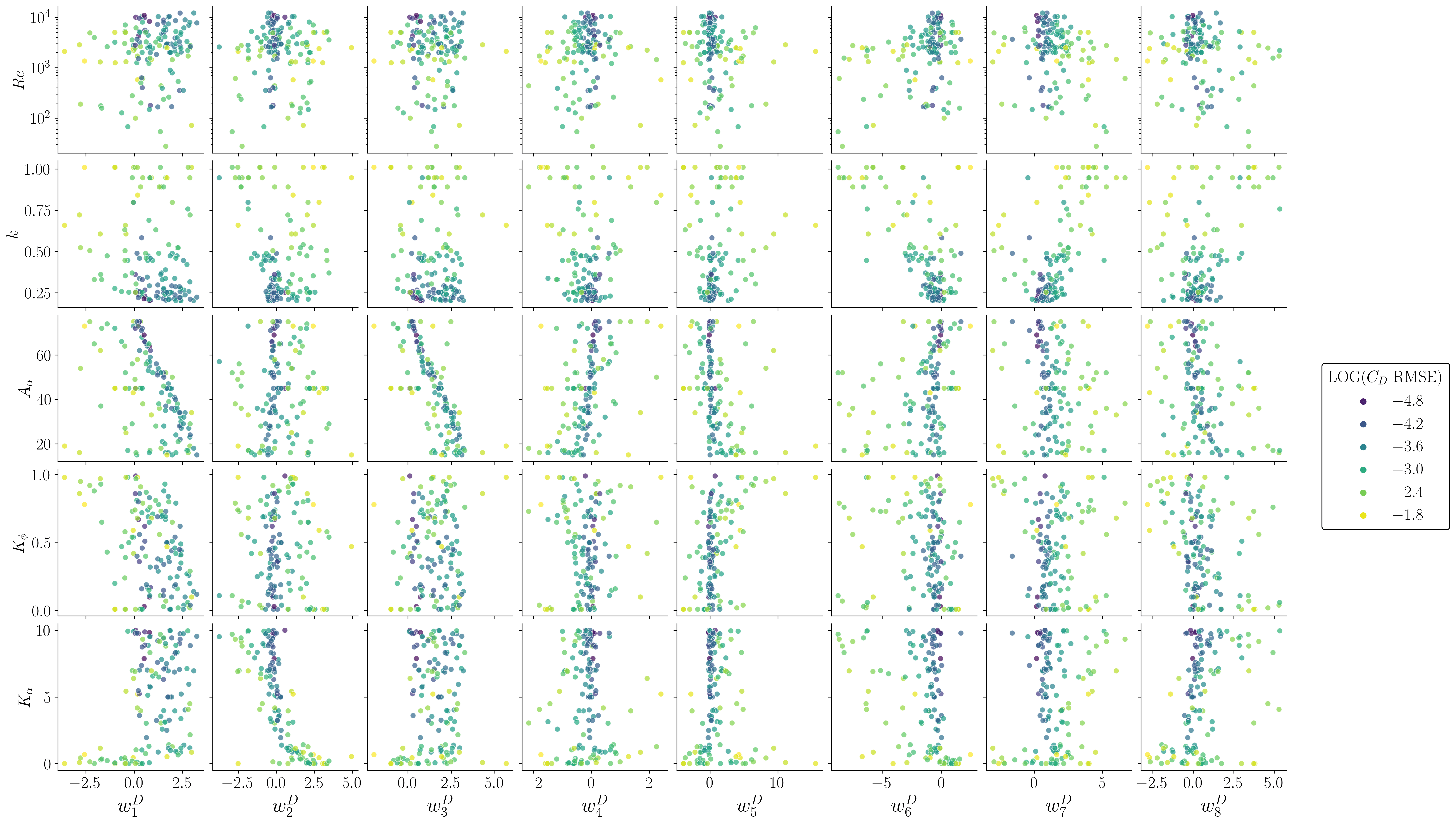}}
		\caption{Lift (a) and Drag (b) IC model weights from the training database vs OOC kinematic parameters. Hue is proportional to the logarithm of RMSE.}
		\label{fig:weights-vs-ooc}
	\end{figure*}

	It is thus interesting to compare these performances with the IC Ridge regression, which always agrees with the data. This shows that the ROM's mispredictions are not due to limits of the in-cycle basis in \eqref{Basis}, but difficulties in regressing the weights in the OOC GP regression. The kinematics in this test case is characterized by sharp transients and large added-mass forces at stroke reversal, producing distinct spikes in lift and drag. Interestingly, the Ridge regression performs satisfactorily because the chosen basis features adapt to this challenging condition. This can be seen in Figure \ref{fig:test-cases-basis}, which illustrates the eight normalized basis functions supporting the regression in the five cases of Figure \ref{fig:test-cases}. In case 5, for example, the bases $b_4(t)=\dot{\alpha}^2$, $b_5(t)=\dot{\phi}\ddot{\phi}$ and $b_7(t)=\sin(2\alpha)\dot{\alpha}$ naturally follow the large spikes in the acceleration and allows the Ridge regression to capture the required sharp gradients in the aerodynamic coefficients.

	The same is true for basis $b_5(t)$ in the third test case. The model capacity of this basis appears abundant for the IC regression, with important redundancies in some cases (e.g. $b_6\approx -b_7$ in case 5). The redundancy produces badly conditioned basis matrices, but the regularization is sufficiently robust to handle all the investigated kinematics. Future development will aim to reduce this redundancy, using a Gram-Schmidt orthogonalization before the IC regression. 
	
	Concerning the limitations in the GP-based OOC regression, figure \ref{fig:test-cases-weight-error} shows the absolute error between the optimal weight computed in the IC regression and the ones predicted by the OOC GP regression. As expected, the largest discrepancies occur for case 3, particularly on the bases $b_2$ and $b_5$. These are primarily involved in the prediction of peak loads at stroke rehearsal. However, especially for $b_5$, the basis is also active during the midstroke. Therefore, a large weight on this basis forces others to compensate. This delicate balance is much less present in the other conditions, explaining why the OOC performs poorly unless more training data is included in this region of the parameter space.
	
	Focusing on training data, the weights from the Ridge regression can be used to analyze the robustness of the regression within the parameter space. Intuitively, a robust Ridge regression is characterized by weights of comparable magnitude, while a large variance is often linked to overfitting problems \cite{Bishop2011}.
	Figure \ref{fig:weights-vs-ooc} shows the optimal weights $\mathbf{w}$ from the IC regression for both lift and drag coefficients as a function of the OOC parameters. The hue of the scatter plot is linked to the RMSE on a logarithmic scale computed from the full model prediction. The scatter plot for the lift coefficient shows that most weights are small far from the boundaries, where the RMSE is also low. On the other hand, the model is less accurate on the drag prediction; the larger RMSE is associated with a larger spreading of the weights, even in the inner portions of the parameter space.

	A pattern is visible for the weights $w_1$ and $w_3$ versus the amplitude of the pitching angle $A_\alpha$ for both coefficients. These are linked to the bases $b_1=\cos(\phi)$ and $b_3=\dot{\phi}^2$, i.e. translation forces. As a result, these terms mostly contribute to time-averaged forces; hence the trends observed in Figure \ref{fig:pairwise-mean}. We can also observe how the model error for $C_L$ increases monotonically with $A_\alpha$, which is associated with a departure from QS assumption.

\begin{figure}[!htb]
		\centering
		\includegraphics[width=0.95\linewidth]{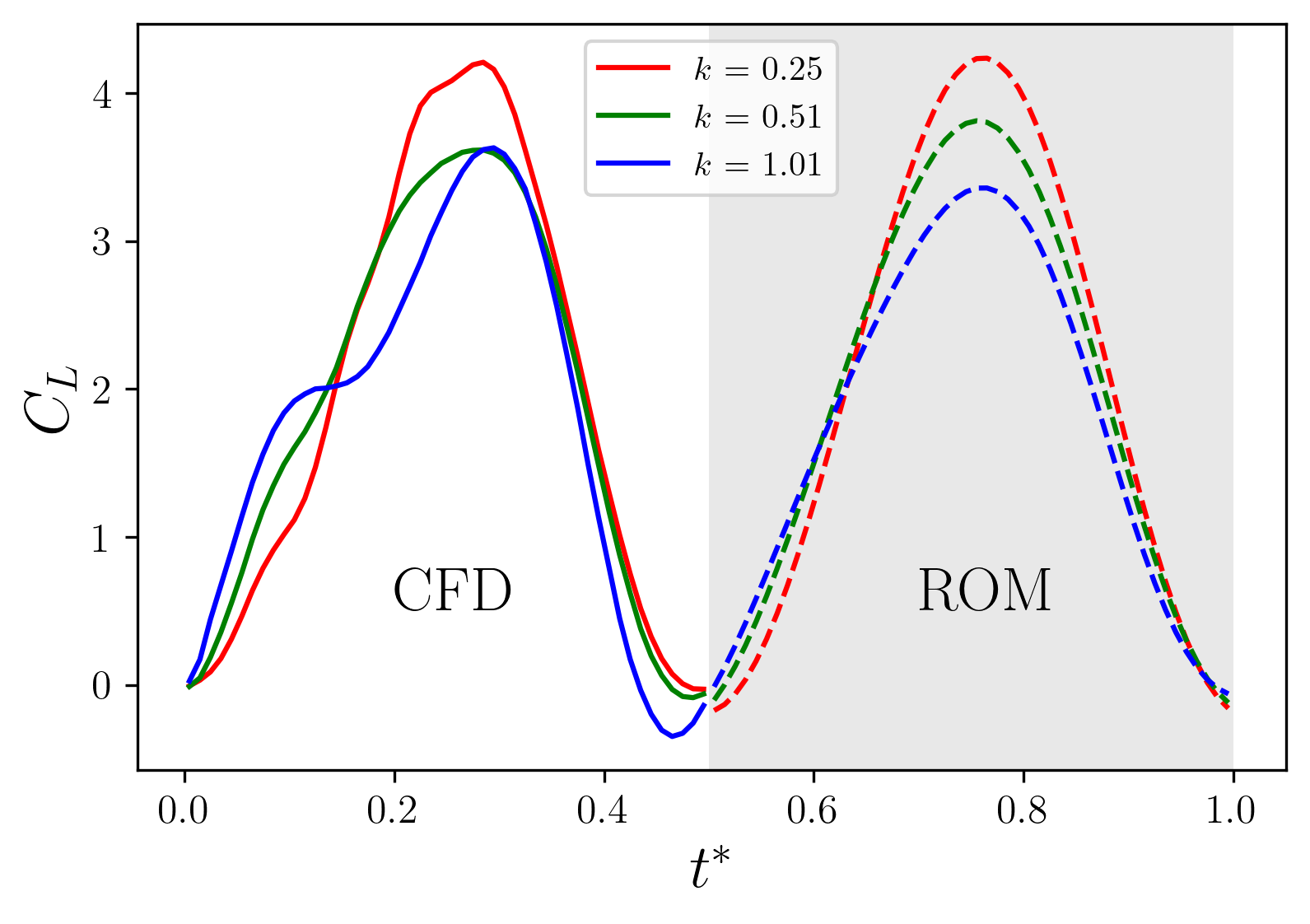}
		\caption{Influence of $k$ on $C_L$ for $Re \approx 5000$,  $A_{\alpha}$=45° and $K_{\phi}=K_{\alpha}=0.01$. CFD results are shown on the first half-cycle (solid lines), and ROM on the second half-cycle (dashed lines).} 
		\label{fig:cl-k}
	\end{figure}

	The parameter that mostly correlates with poor predictions and weight spreading is the reduced frequency $k$, with the worst performances obtained at the largest values. To illustrate the impact of this parameter on the lift coefficient, Figure \ref{fig:cl-k} shows the $C_L$ profile for three cases with distinct $k$ keeping other parameters fixed. These test cases are characterized by $k=0.25$, $k=0.51$ and $k=1.01$, with $Re\approx5000$, $A_{\alpha}$=45° and $K_{\phi}\approx K_{\alpha}\approx 0.01$. These kinematics produce smooth harmonic motion for flapping and pitching. However, a higher $k$ (lower $A_{\phi}$) for the same $Re$ imposes a higher flapping frequency, thus higher velocity/acceleration. The first half cycle of the plot shows the CFD results, and the second half shows the ROM prediction. Doubling $k$,  $\langle C_L \rangle$ decreases 3\% from 1.97 to 1.91, and although the peak value is lower, a faster rise is produced during the initial stroke. One could expect stronger interactions with the wake for a smaller flapping amplitude. This was also observed in the $C_L$ history from CFD, where higher $k$ showed a larger discrepancy between the first cycle (no wake) and subsequent ones. In addition, higher accelerations create greater added mass effects. On the other hand, there is less time/span for the LEV formation, which explains the drop in peak lift. The last case for $k$ equal to 1 exhibits the latter trends more explicitly, with an even greater rise at the start of each stroke and periods of negative lift at the end. An inflection point is visible after the initial rise, possibly due to the transition between wake interaction and LEV mechanisms. The net effect of the wake interaction and LEV for this case is more detrimental, compared to $k=0.5$, with $\langle C_L \rangle$ = 1.77. The wake interaction is linked to induced downwash (reducing the effective angle of attack) and effects on wing tip vortices, changing the pressure distribution around the wing and reducing lift \cite{Sun2002,Nakata2015}. For these cases with variable $k$ but moderate pitching amplitude, the ROM can capture the trends in peak lift but tends to smooth the profiles.
	
\begin{figure*}[!ht]
		\centering
		\subfloat[Wing motion angles.]{\label{fig:good-IC-ridge-angles}\includegraphics[width=0.4\linewidth]{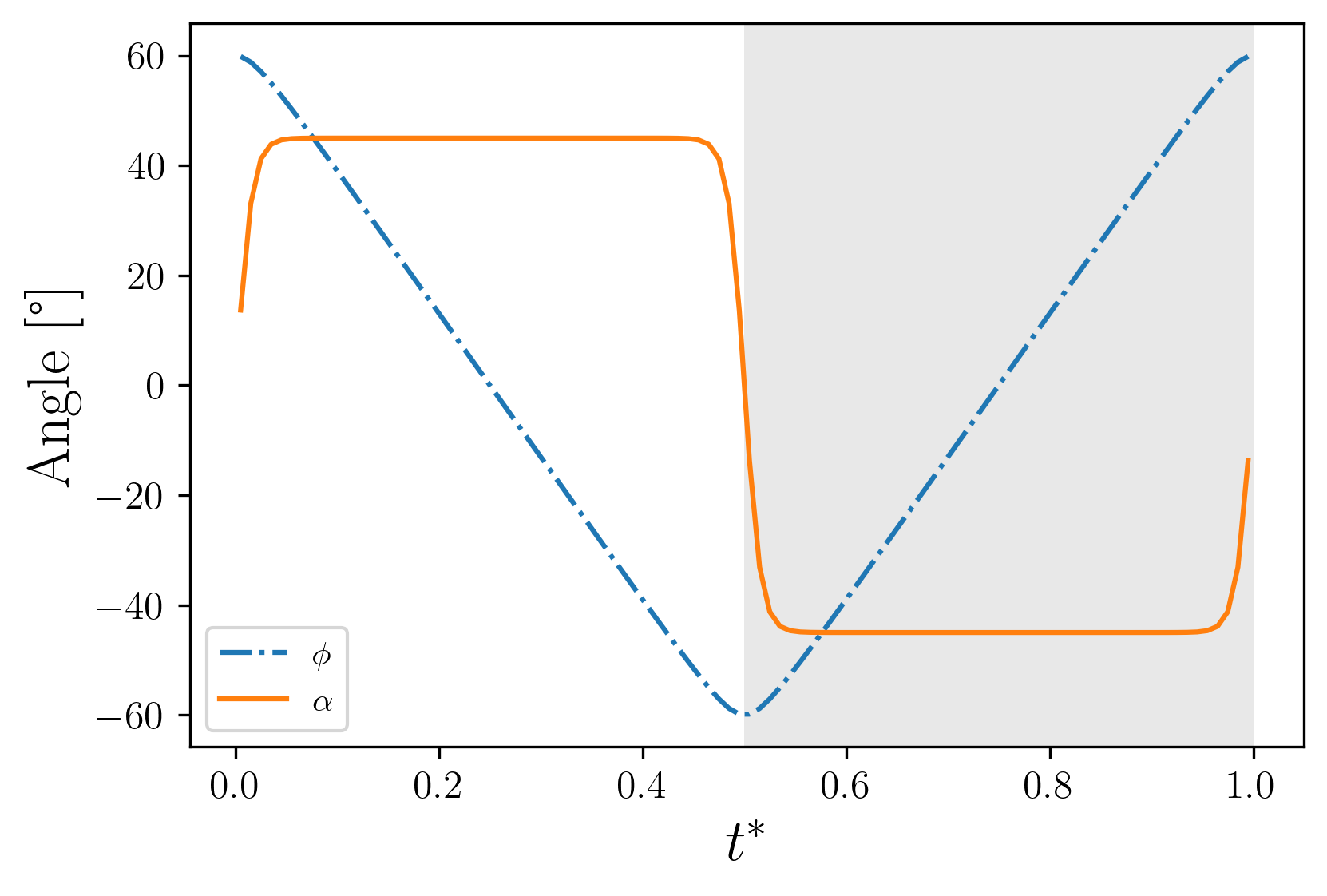}} \quad
		\subfloat[$C_L$ history from CFD (solid line) and ROM (dashed line) with horizontal lines representing $\langle C_L \rangle$.]{\label{fig:good-IC-ridge-CL}\includegraphics[width=0.4\linewidth]{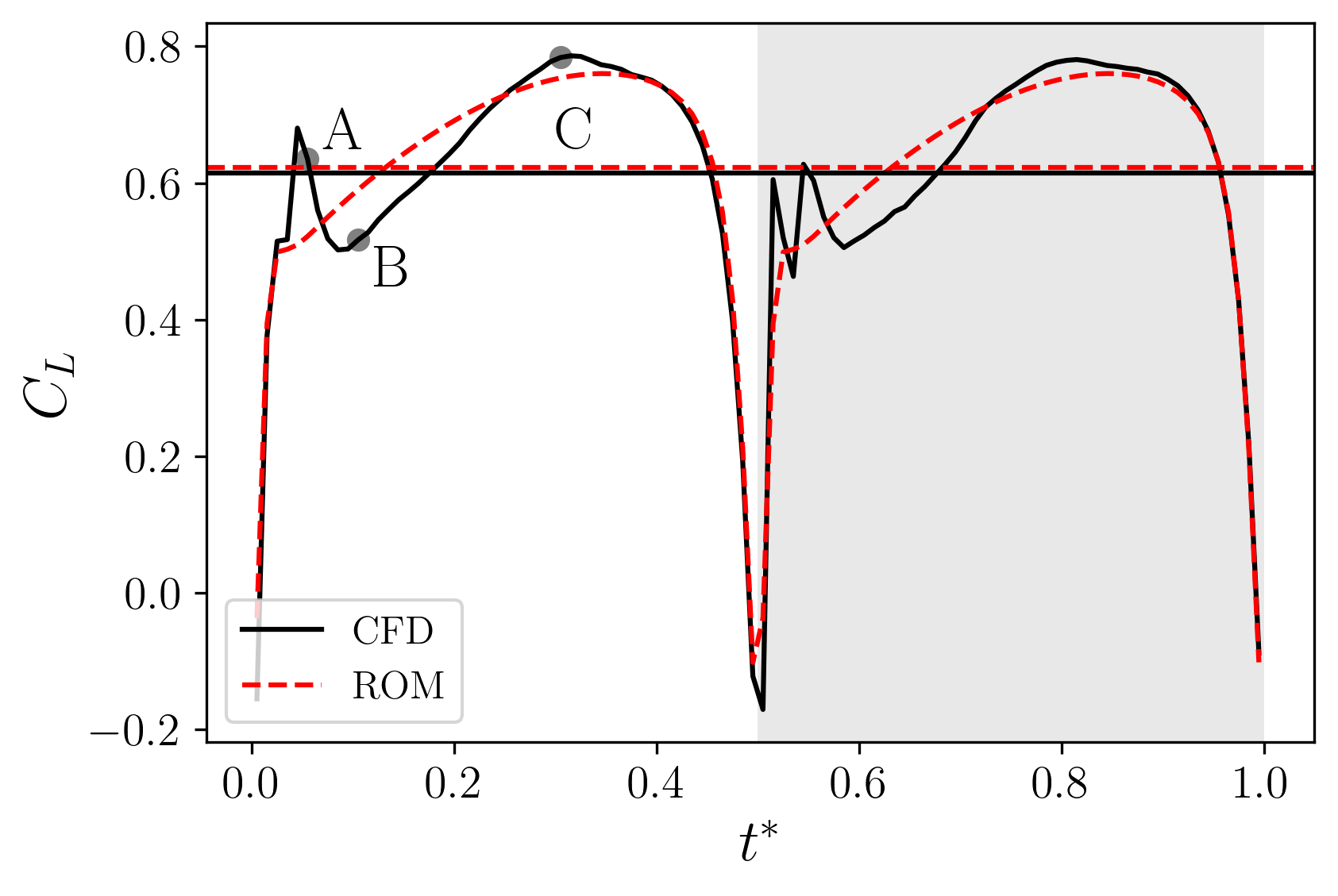}} \\      
		\subfloat[CFD flow visualization. Viewpoint is top-down, showing upper wing surface. Top row: vortical structures defined from $Q$ criterion, coloured by helicity magnitude (wing shown in solid green). Bottom row: pressure coefficient contours.]{\label{fig:dynamic}\includegraphics[width=0.8\linewidth]{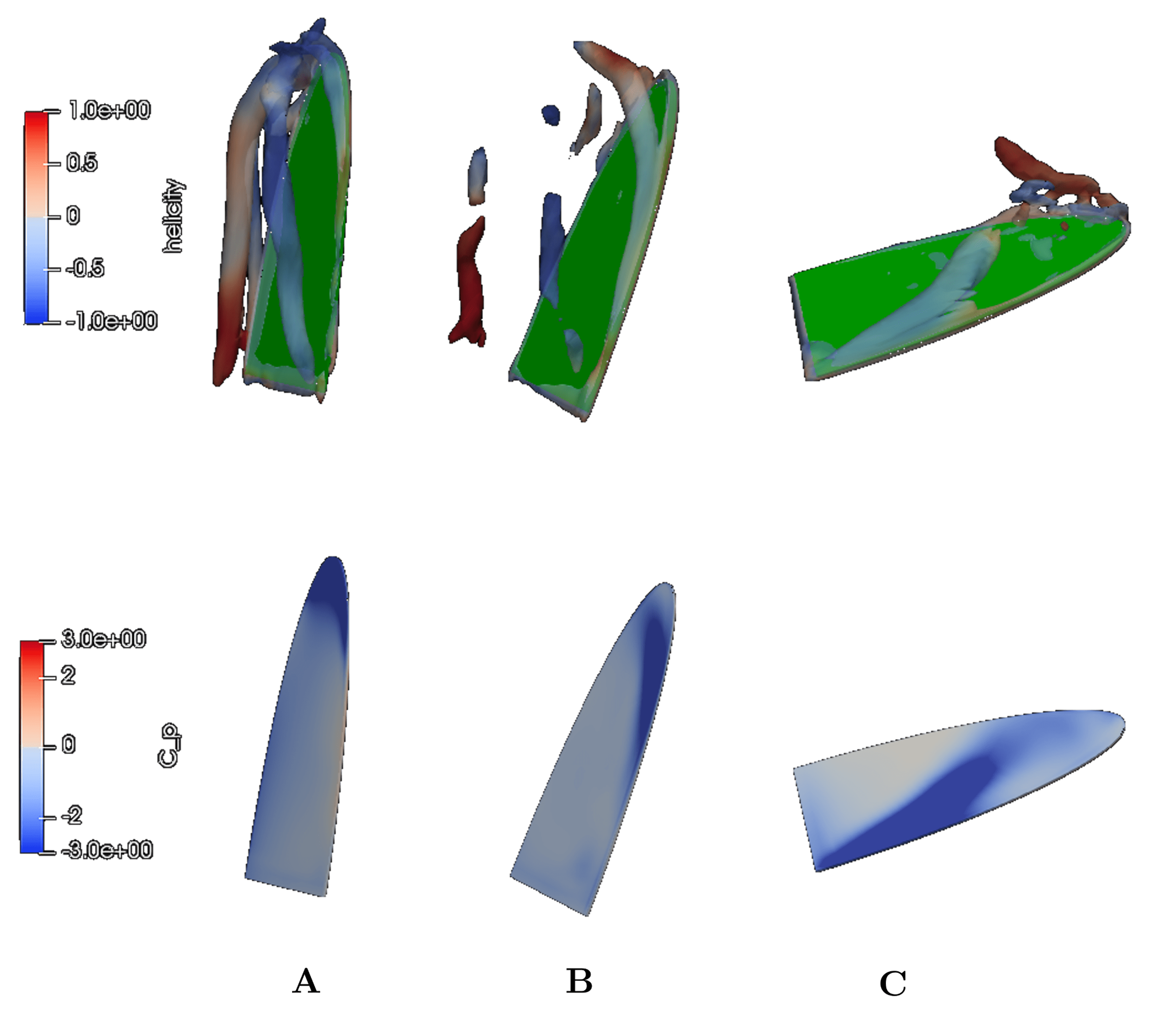}}
		\caption{Training case at limit $Re$ and shape factors.  $A_{\phi} = 60$° ($k=0.25$), $A_{\alpha} = 45$°, $K_{\phi}$ = 0.99, $K_{\alpha}$ = 10, $Re = 10^4$. Wing motion angles in (a). $C_L$ history and time average is shown in (b). CFD flow visualization snapshots in (c).} 
		\label{fig:good-IC}
	\end{figure*}

\begin{figure*}[!ht]
		\centering
		\subfloat[Wing motion angles.]{\label{fig:worse-IC-ridge-angles}\includegraphics[width=0.4\linewidth]{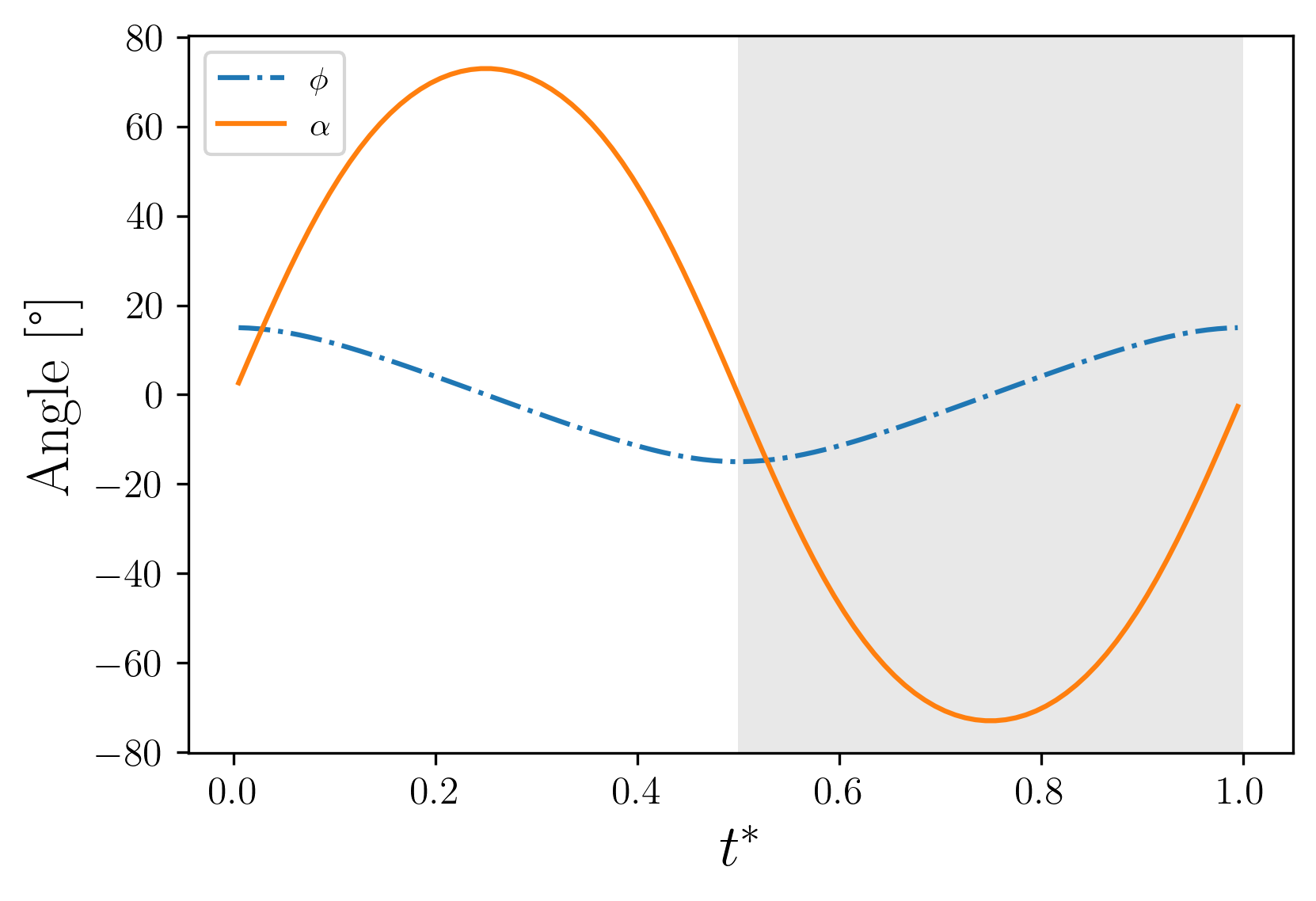}} \quad
		\subfloat[$C_L$ history from CFD (solid line) and ROM (dashed line) with horizontal lines representing $\langle C_L \rangle$.]{\label{fig:worse-IC-ridge-CL}\includegraphics[width=0.4\linewidth]{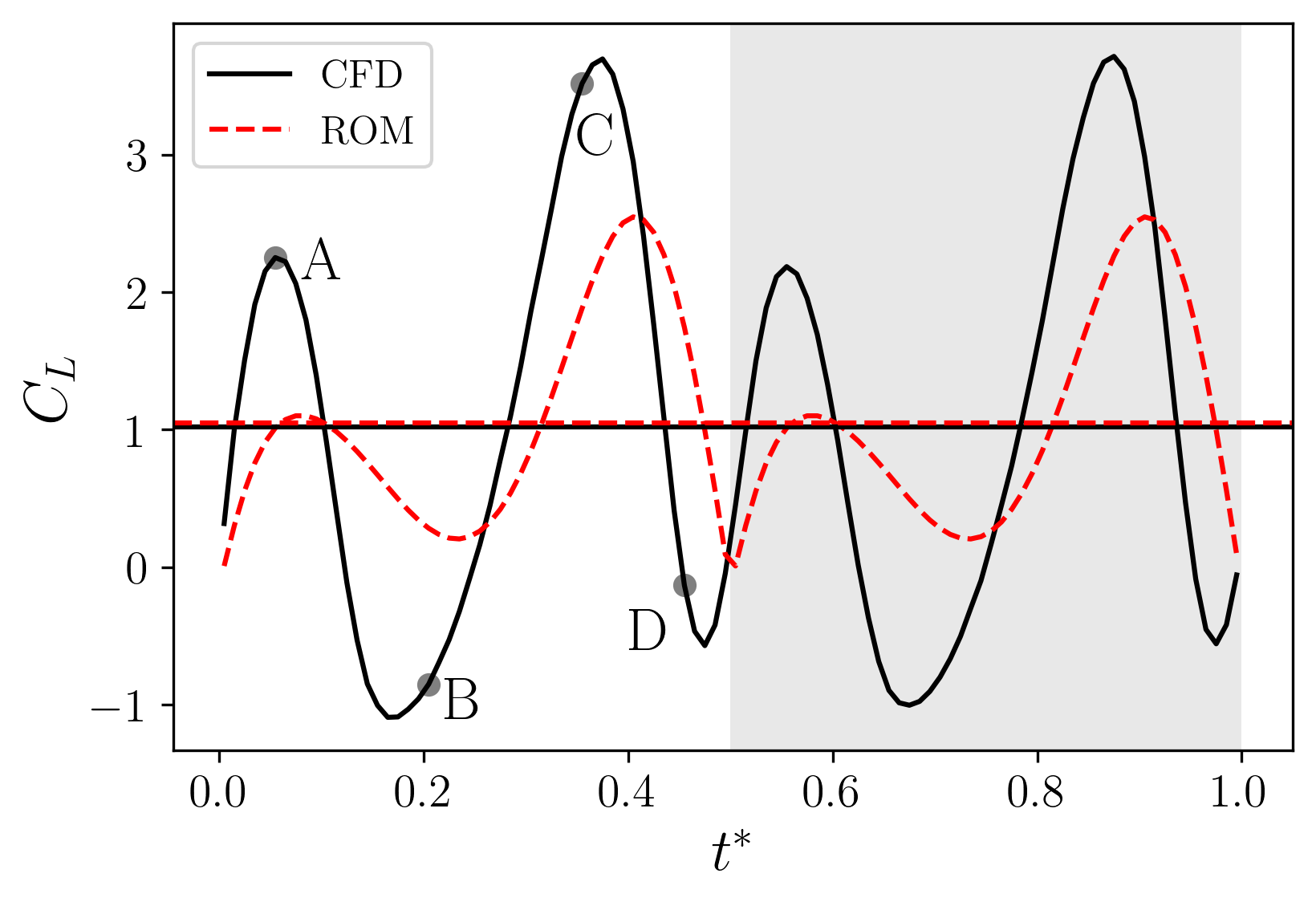}} \\      
		\subfloat[CFD flow visualization. Viewpoint is top-down, showing the upper wing surface. Top row: vortical structures defined from $Q$ criterion, coloured by helicity magnitude (wing shown in solid green). Bottom row: pressure coefficient contours.]{\label{fig:unstable-lev}\includegraphics[width=0.8\linewidth]{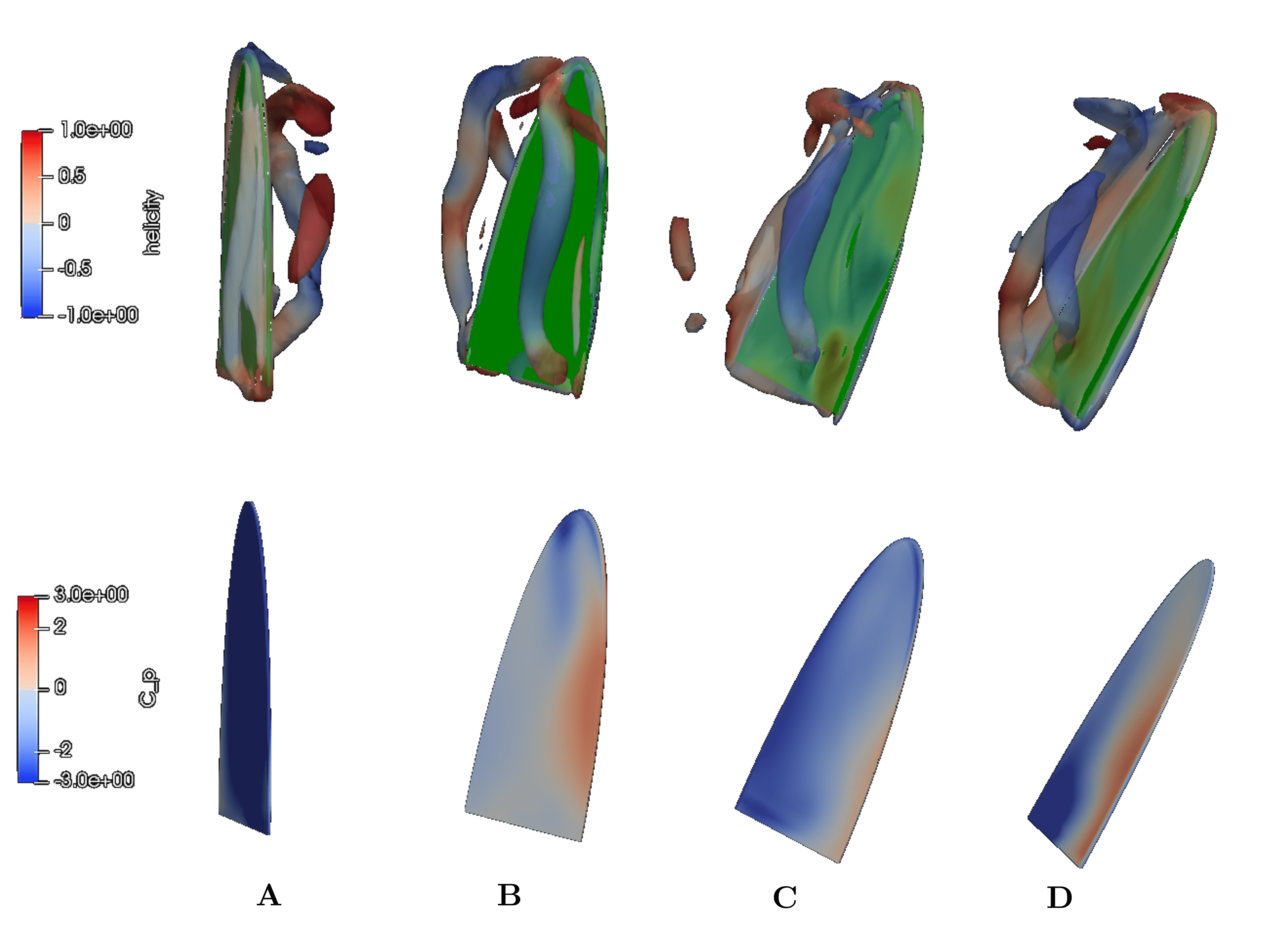}}
		\caption{Training case with highest RMSE for $C_L$.  $A_{\phi} = 15$° ($k=1.01$), $A_{\alpha} = 73$°, $K_{\phi}$ = 0.78, $K_{\alpha}$ = 0.68, $Re = 1355$. Wing motion angles in (a). $C_L$ history and time average are shown in (b). CFD flow visualization snapshots in (c).} 
		\label{fig:worse-IC}
	\end{figure*}

	Finally, we conclude this section with an insight on the flow dynamics for two representative test cases: one with extreme dynamics in terms of $Re$ and shape factors with very low error (RMSE=0.005) and the one that with highest (RMSE=0.11). These are analyzed in Figures \ref{fig:good-IC} and \ref{fig:worse-IC}, respectively.
	
	In each figure, subfigure (a) shows the underlying flapping kinematics (with parameters recalled in the caption), subfigure (b) shows the evolution of the lift coefficient from the CFD together with the ROM prediction while subfigure (c) provides a flow visualization (first row) and pressure distribution on the wing (second row) for different snapshots (also labelled in subfigure b). In the flow visualization, vortical structures via isosurface of Q field \cite{vortex-identification}, with colour contour map in terms of normalized helicity $h$ similar to \citet{Bos2013}. These quantities are defined as 
	
	\begin{equation}
		\label{eq:Q}
		Q = \frac{1}{2}(||\bm{\Omega}||_F^2-||\mathbf{S}||_F^2) \,\, \mbox{and}\,\,
		h = \frac{\mathbf{v}\cdot \mathbf{\omega}}{||\mathbf{v} || \, ||\mathbf{\omega}||}\,,
	\end{equation} \, where $\mathbf{v}=(u,v,w)$ is the velocity vector, 
	$\mathbf{S} = \frac{1}{2}(\nabla \mathbf{v}+\nabla \mathbf{v}^\top)$ and $\bm{\Omega} = \frac{1}{2}(\nabla \mathbf{v}-\nabla \mathbf{v}^\top)$ are the symmetric and anti-symmetric parts of the velocity gradient tensor, $||\bullet||_F$ is the Frobenius norm of a matrix and $\cdot$ denotes scalar product. Positive values of Q indicate that vorticity exceeds strain while helicity measures the alignment of velocity and vorticity.
	
	Focusing on the test case with the low RMSE (Figure \ref{fig:good-IC}, characterized by sharp kinematics at the half-stroke, where added mass forces are the highest: at $t^*=0.05$, when the pitching angle has reached its maximum value $\alpha=$ 45°, a sharp peak is observed, followed by a sudden drop. 
	The visualizations and pressure contours show that the peak in the lift (snapshot A) coincides with the presence of large vortical structures detaching from the leading edge and the trailing edge (from the previous stroke reversal) while the sudden drop (snapshot B) occurs when the LEV detaches. The largest lift occurs at $t^*=0.35$ (instant C), when a LEV has re-established and remains attached while moving towards the root of the wing, where it creates a large suction area.
	
	Although the proposed ROM misses the peaks and drops in snapshots A and B, the overall trends are well captured. This result is remarkable because this flapping kinematics is outside the range of validity of classic quasi-steady formulations.
	
	Finally, focusing on the case with the largest RMSE (Figure \ref{fig:worse-IC}), this is characterized by moderate shape factors kinematics and a comparatively small Reynolds number but large pitching angles and small flapping amplitudes.
	The kinematics trigger a highly unsteady phenomenon which produces large fluctuations in the aerodynamic forces with both positive and negative peaks. The first positive peak occurs at $t^*=0.2$ when the flapping acceleration is the highest and hence the added mass contribution. 
	A large vortex from the previous stroke is also present underneath the wing (wing-wake interaction; see snapshot A) and tends to decrease lift. 
	The second positive peak occurs at $t^*=0.3$, when a vortex sheet (snapshot C) composed of both LEV and TEV is formed on the upper side.
	The negative peaks (snapshots B and D) are both associated with over-pressures near the leading edge. At those instants, the LEV and TEV are both detached, and the wake of the previous stroke induces an impingement flow. The strongest wing-wake interaction is then visible for a high value of $k$ as discussed in Fig. \ref{fig:cl-k}.

	\section{Conclusions}\label{sec:conclusions}
	This work proposes a robust data-driven QS ROM to predict instantaneous lift and drag in flapping ellipsoid rigid wings in hovering conditions. The model was trained and tested on an extensive CFD database of 165 simulations using the overset method. The database covers a broad range of Reynolds numbers ($10^2-10^4$) and flapping/pitching amplitudes ($15^{\circ}-75^{\circ}$). 
	
	The data-driven ROM was constructed as a combination of Ridge regression (IC regression), leveraging a basis of nonlinear kinematic features, and Gaussian Processes (OOC regression) to adapt to various flapping kinematics. The Gaussian Process also allows estimating model uncertainties at each prediction. Moreover, the proposed ROM solely requires the kinematic parameters as input and does not rely on the spanwise discretization of forces and velocities.
	
	The CFD dataset was extensively explored to assess the parameter space's sampling uniformity and identify trends and regions of optimal $C_L/C_D$ flapping kinematics. The proposed ROM achieved good performance (with a PCC of 0.93 on test data). Moreover, the best performances are achieved in the region near the optimal lift-to-drag ratio, where the RMSE is found to be of the order of 1\%. 
	
	A detailed analysis of the model performance shows that the main limitation is in the OOC regression, while the IC regression performs remarkably well, even in particularly aggressive kinematics. A more extensive dataset, combined with adaptive kernels for the Gaussian Process, could offer further improvements. Nevertheless, successful ROM performances are particularly relevant, considering that many of the near-optimal investigated conditions are characterized by unsteady mechanisms (revealed via CFD visualizations) that usually fall well beyond the reach of the QS formalism. 
	
	Future work will aim at extending the proposed approach to more complex flapping conditions (e.g. adding wing flexibility and whole-body configurations). On the uncertainty evaluation side, more sophisticated heteroscedastic models can also be considered. 
	
	In conclusion, the success of the presented ROM highlights the potential of data-driven methods to provide generalizable models and stretch the validity of the QS formulation. Furthermore, such fast and reliable ROMs could enable model predictive control of FWMAVs, where extremely fast dynamics require timely predictions of the aerodynamic forces acting on the wings. Continued research on these ROMs should promote the development and use of engineered FWMAVs in their different applications.
	
	\begin{acknowledgments}
		This work was carried out in the framework of the first author's Research Master program at the von Karman Institute for Fluid Dynamics and was supported by a Fellowship of the Belgian American Educational Foundation (BAEF). R. Poletti is supported by Fonds Wetenschappelijk Onderzoek (FWO), Project No. 1SD7823N.
	\end{acknowledgments}
	
	\section*{Data Availability Statement}
	The data that support the findings of this study are available from the corresponding author upon reasonable request.
	
	%\nocite{*}
	
	\section{References}
	\bibliography{Calado_et_al_2022}% Produces the bibliography via BibTeX.
	
\end{document}